\documentclass[aps,prd,twocolumn,superscriptaddress]{revtex4-1}
\usepackage{epsfig}
\usepackage{graphicx}

\usepackage{verbatim} %my own add...to use comments
\usepackage{amsmath}

\global\long\def\m#1{\mbox{$#1$}}

\begin{document}

\title{%
CMB spectral distortions from the decay of causally generated magnetic fields}
% Force line breaks with \\

\author{Jacques M. Wagstaff}\email{jwagstaff@hs.uni-hamburg.de}
\author{Robi Banerjee}
\affiliation{Hamburger Sternwarte, Gojenbergsweg 112, 21029 Hamburg, Germany}
\affiliation{Nordita\\
KTH Royal Institute of Technology and Stockholm University\\
Roslagstullsbacken 23, SE-106 91 Stockholm, Sweden}

\date{\today}

\begin{abstract}
We improve previous calculations of the CMB spectral distortions due to the decay of primordial magnetic fields. We focus our studies on causally generated magnetic fields at the electroweak and QCD phase transitions. We also consider the decay of helical magnetic fields. We show that the decay of non-helical magnetic fields generated at either the electroweak or QCD scale produce $\mu$ and $y$-type distortions below $10^{-8}$ which are probably not detectable by a future PIXIE-like experiment. We show that magnetic fields generated at the electroweak scale must have a helicity fraction $f_*>10^{-4}$ in order to produce detectable $\mu$-type distortions. Hence a positive detection coming from the decay of magnetic fields would rule out non-helical primordial magnetic fields and provide a lower bound on the magnetic helicity. 
\end{abstract}

%\pacs{98.80.Cq}
\maketitle

%\begin{widetext}

%%---------------------------------------------------------
%%---------------------------------------------------------

\section{Introduction}

Magnetic fields are observed throughout the cosmos; from galaxies at high and low redshifts \cite{BFields_galax,BFields_Bernet,greenpeas}, in galaxy (super)clusters \cite{BFields-Clusters,BFields_SClusters}, and in the voids of the large scale structure \cite{Neronov:1900zz}.
Although still under debate, the difficulties of astrophysical mechanisms in explaining such observations promote the idea of a primordial origin for magnetic fields i.e. fields generated in the early Universe before structure formation. There are a number of theoretical mechanisms proposed to generate primordial magnetic fields, such mechanisms for example involve inflation \cite{Turner:PMF}, first-order phase transitions \cite{B-phaseT}, or vorticity generation \cite{Harrison,Dolgov:2001nv}.

If primordial magnetic fields were indeed generated, they would suffer decay on small scales due to magnetohydrodynamics (MHD) effects \cite{Christensson:2000sp,Banerjee:2004df,Campanelli:2013iaa}. The dissipation of magnetic fields injects energy into the plasma, if this occurs in the early Universe when the CMB is being formed \cite{CMB_BB}, distortions to its black-body spectrum can be generated \cite{Sunyaev1969,HuSilk:1993,Khatri:2012rt,Chluba:2012gq,Pajer:2013oca,KK14,Jedamzik_CMBdist,Miyamoto:2013oua,Amin:2014ada,Tashiro:2014pga}. Distortions to the CMB black-body spectrum come in different types depending on the epoch of energy injection. At relatively early times a $\mu$-type distortion can be generated, where the photon Bose-Einstein distribution develops a non-vanishing chemical potential. Whereas at later times a Compton $y$-parameter can be generated giving a $y$-type distortion. Mixed distortions are also possible \cite{Khatri:2012tw}. Any mechanism which injects energy into the plasma in the early Universe has the potential to generate such distortions. Other mechanisms in the early Universe include the dissipation of primordial acoustic waves \cite{Hu:1994bz,Chluba:2012gq}, decaying or annihilating relic particles \cite{McDonald:2000bk,Chluba:2011hw,Chluba:2013wsa}, and the evaporation of primordial black holes or cosmic strings \cite{Carr:2009jm,Tashiro:2012pp}.

In this paper we consider magnetic fields generated by some causal process in the early Universe. In particular we consider the possibility of magnetic fields generated by first-order phase transitions at either the electroweak (EW) or the quantum chromodynamics (QCD) scale \cite{B-phaseT}. The spectrum on large scales for magnetic fields generated in such a process is highly constrained due to causality reasons \cite{CausalBfield}. The slope of the large scale spectrum also determines the subsequent MHD evolution of magnetic energy and coherence scale \cite{Banerjee:2004df,Campanelli:2007tc,Campanelli:2013iaa}. As the small scales dissipate into heat in a turbulent plasma, the peak of the spectrum moves down along the large scale spectrum \cite{Banerjee:2004df,Saveliev:2012ea,Saveliev:2013uva,Campanelli:2013iaa}. 
The amplitude of magnetic helicity also determines the evolution of magnetic fields. Helicity conservation in the early Universe slows down the decay rate of fully helical fields compared to non-helical fields and leads to an inverse cascade of energy from small scales to large scales \cite{Christensson:2000sp,Banerjee:2004df}. Hence causally generated magnetic fields and their initial helicity fraction can be constrained by CMB spectral distortions.

In Section \ref{sec:mag_decay} we use the results of Refs.~\cite{Banerjee:2004df,Campanelli:2007tc,Campanelli:2013iaa} to determine the decay rates of magnetic fields as functions of their initial strength, coherence length and initial helicity fraction. From these decay rates, in Section \ref{sec:general}, we calculate the CMB spectral distortions generated and analyse the parameter space. We conclude in the final Section.

%In section \ref{sec:general} we summarise known results for the CMB spectral %distortions given some energy injection. In section

%%---------------------------------------------------------
%%---------------------------------------------------------

\section{The decay of magnetic fields in the early Universe}\label{sec:mag_decay}

In order to calculate the CMB spectral distortions we need to know the energy injected into the primordial plasma. For decaying magnetic fields, the energy injection rate is given by \cite{KK14,Jedamzik_CMBdist}
\begin{equation}
\frac{\textrm{d}Q}{\textrm{d}t}\equiv
-a^{-4}\frac{\textrm{d}\tilde{\rho}_B}{\textrm{d}t}\,,
\end{equation}
where $\tilde{\rho}_B$ is the comoving energy density i.e. \m{\tilde{\rho}_B\equiv a^4\rho_B}. The averaged magnetic energy density is obtained by integrating over the \emph{local} energy density 
\m{u_B=\tilde{\mathbf{B}}^2/8\pi}, here \m{\tilde{\mathbf{B}}\equiv a^2\mathbf{B}} is the comoving magnetic field, 
\begin{equation}\label{eq:epsilon_B}
 \tilde{\rho}_B=\frac{1}{V}\int u_B\mathrm{d}\mathbf{r}
  =\frac{1}{8\pi}\int|\tilde{\mathbf{B}}(\mathbf{k})|^2\mathrm{d}\mathbf{k}
  \equiv \tilde{\rho}\int M_k\mathrm{d}k
  \,,
\end{equation}
where \m{\tilde{\rho}} is the total comoving energy density and $M_k$ is the magnetic spectral energy. Here we emphasise that the spectrum on large scales, parametrized by \m{M_k\propto k^{n+2}}, for causally generated magnetic fields is constrained by \m{n\geq2} \cite{CausalBfield}. The most shallow, and expected, scaling \m{n=2} is also confirmed by numerical simulations giving \m{M_k\propto k^{4}} \cite{Saveliev:2012ea,Brandenburg:2014mwa}. Assuming that the magnetic energy is concentrated at the \emph{Integral} scale (index $I$), which defines the peak of the spectrum in Fourier space, we can write
\m{\tilde{\rho}_B=\tilde{\rho}\int k M_k\mathrm{d}\ln k\simeq
\tilde{\rho} k_I M_I}, adopting the conventions of Ref.~\cite{Saveliev:2012ea}
where the wave vector $k$ is also comoving. Since the photon energy density scales as \m{\rho_\gamma\propto a^{-4}}, we can write
\begin{equation}\label{eq:dQdz}
\rho_{\gamma}^{-1}\frac{\textrm{d}Q}{\textrm{d}z}=
-\rho_{\gamma,0}^{-1}\frac{\textrm{d}\tilde{\rho}_B}{\textrm{d}z}
\simeq-\rho_{\gamma,0}^{-1}\frac{\textrm{d}\left(\tilde{\rho} k_I M_I\right)}{\textrm{d}z}\,,
\end{equation}
where \m{a_0\equiv a(T_0)=1} today.

The magnetic field strength $B_I$ and coherence length $L_I$ (identified as $2\pi/k_I$) evolve during the radiation dominated era due to turbulent MHD effects \cite{Banerjee:2004df,Durrer:2013pga,Campanelli_free-turb,Wagstaff:2013yna}. In the following we quote the results from the detailed analytical and numerical analysis in Refs.~\cite{Banerjee:2004df,Campanelli:2007tc,Campanelli:2013iaa}.
The evolution of the field strength and coherence length depends on the state of the plasma and on the properties of the magnetic field, in particular its helicity.
The average helicity density is given by
\begin{equation}
 h_B=\frac{1}{V}\int\left(\mathbf{A}\cdot\mathbf{B}\right)\mathrm{d}\mathbf{r}
  =\rho\int\mathcal{H}_k\mathrm{d}k
  \,,
\end{equation}
where \m{\mathbf{B}=\nabla\times\mathbf{A}}. For a magnetic field with helicity, on any given scale $k$ there is a \emph{realizability} condition given by \m{|\mathcal H_k|\leq 8\pi M_k/k}, where the helical spectrum $\mathcal H_k$ is defined following the conventions of Ref.~\cite{Saveliev:2013uva}. From the above we can define \m{f\equiv k\mathcal H_k/8\pi M_k} as the helicity fraction,
where $f=0$ for the non-helical case and $f=1$ for the maximally helical case.
The helicity density is a useful quantity since it is conserved in the early Universe \m{h_B\simeq\mathrm{const.}} when the conductivity \m{\sigma=1/4\pi\eta\rightarrow\infty} \cite{Biskamp1993}, and the conservation of magnetic helicity determines the evolution of the magnetic field strength and coherence length.

In the turbulent regime, where kinetic Reynolds numbers are large \m{R_e\gg1}, the general decay law for the magnetic energy is \m{M_I\propto a^{-2(n+2)/(n+5)}} and for the comoving integral scale \m{L_I\propto a^{2/(n+5)}} \cite{Banerjee:2004df,Saveliev:2012ea,Saveliev:2013uva}, hence \m{\tilde{B}_I\propto a^{-5/7}}, \m{k_I\propto a^{-2/7}} and \m{\tilde{\rho}_B\propto a^{-10/7}},
%
\begin{comment}
%
\begin{equation}\label{eq:BL_nonHelical}
 \tilde{B}_I\propto a^{-\frac57}\,,\qquad
k_I\propto a^{-\frac27}
\qquad\textrm{and}\qquad
\tilde{\rho}_B\propto a^{-\frac{10}{7}}\,,\qquad\text{(turbulent damping, non-helical)}
\end{equation} 
%
\end{comment}
%
where \m{n=2} is used due to causality constraints for the large scale spectrum~\cite{CausalBfield,Banerjee:2004df,Saveliev:2012ea,Saveliev:2013uva}. 
These decay laws for non-helical magnetic fields, where the helicity fraction \m{f\ll1} if not zero, are obtained through analytical considerations in Refs.~\cite{Banerjee:2004df,Campanelli:2007tc,Campanelli:2013iaa} and confirmed numerically in Ref.~\cite{Banerjee:2004df}.
If the magnetic field has non-zero helicity, the helicity density in this case would grow as \m{\mathcal H_I\propto a^{2/7}}, and eventually the magnetic field would become fully helical \m{f=1}. The conservation of magnetic helicity ensures the relation \m{\tilde{\rho}_B\propto \tilde{B}^2_I\propto k_I}. The decay of magnetic fields slows down in the fully helical case and the decay rate becomes independent of the shape of the large scale spectrum $n$. For the maximally helical case we find \m{\tilde{B}_I\propto a^{-1/3}}, \m{k_I\propto a^{-2/3}} and \m{\tilde{\rho}_B\propto a^{-2/3}}.
%
\begin{comment}
\begin{equation}\label{eq:BL_Helical}
 \tilde{B}_I\propto a^{-\frac13}\,,\qquad
k_I\propto a^{-\frac23}
\qquad\textrm{and}\qquad
\tilde{\rho}_B\propto a^{-\frac23}\,,\qquad\text{(turbulent damping, max. helical)}
\end{equation} 
%
\end{comment}
We can summarise the above turbulent damping (TD) decay laws by
\begin{equation}\label{eq:turb_decay}
(\mathrm{TD}):\qquad
\tilde{\rho}_B\propto a^{-2\frac{p+5}{p+7}}\,,\quad
L_I\propto a^{\frac{2}{p+7}}\,,
\end{equation} 
%
%
\begin{comment}
\begin{equation}\label{eq:turb_decay}
 \tilde{B}_I\propto a^{-\frac{p+5}{p+7}}\,,\qquad
L_I\propto a^{\frac{2}{p+7}}
\qquad\textrm{and}\qquad
\tilde{\rho}_B\propto a^{-2\frac{p+5}{p+7}}\,,\qquad\text{(turbulent damping)}
\end{equation} 
\end{comment}
%
where \m{p=0} for non-helical fields and \m{p=-4} for maximally helical fields.

Here we note that an apparent “inverse transfer” of magnetic energy has been numerically observed for the non-helical case. This effect leads to a weaker evolution for non-helical magnetic fields in the turbulent regime
\m{\tilde{B}_I\propto a^{-1/2}}, \m{k_I\propto a^{-1/2}} and \m{\tilde{\rho}_B\propto a^{-1}}, i.e. \m{p=-3} above, giving \m{B_I\propto L_I^{-1}} \cite{Kahniashvili:2012uj} (see also Ref.~\cite{Campanelli:2004wm}). This is potentially a very interesting and exciting new development in turbulent MHD. However this is a numerically observed effect
under the conditions of high resolutions, and magnetically dominant turbulence~\cite{Brandenburg:2014mwa}. The condition of magnetically dominant turbulence is perhaps not satisfied in the early Universe. Magnetogenesis at first order phase transitions typically produce a lot of turbulent kinetic energy. The generated magnetic field, through dynamo action, comes into equipartition with the kinetic energy, but is unlikely to dominate over the kinetic energy, see for example Ref.~\cite{B-phaseT}. Furthermore, in the study by Ref.~\cite{Brandenburg:2014mwa} it seems that the inverse transfer is less efficient for large Prandtl numbers, but the Prandtl numbers in the early Universe are huge. In any case we will also investigate the CMB spectral distortions considering these scalings to complete the study.

In the viscous regime, where \m{R_e<1}, during particle diffusion, the rapidly growing particle mean free path means that the dissipative time scale is increasing faster than the Hubble time $H^{-1}$. This prevents further dissipation of magnetic energy, and so the magnetic energy freezes-out and remains constant~\cite{Banerjee:2004df}
%
\begin{comment}
\begin{equation}\label{eq:visc_freez}
 \tilde{B}_I\simeq \textrm{const.}\,,\qquad
L_I\simeq \textrm{const.}
\qquad\textrm{and}\qquad
\tilde{\rho}_B\simeq \textrm{const.}\,,\qquad\text{(viscous freezing)}
\end{equation}
\end{comment}
%
\begin{equation}\label{eq:visc_freez}
(\mathrm{VF}):\qquad
\tilde{\rho}_B\simeq \textrm{const.}\,,\quad
L_I\simeq \textrm{const.}\,,
\end{equation}
The above evolution in the viscous freezing (VF) stage occurs regardless of magnetic helicity. However, as the mean free path increases further particles begin to free-stream out of overdensities. In this case the interaction between the fluid and the background is described by a drag force, which is a decreasing function of time. This leads to a situation where the magnetic dissipative time scale can become smaller than the Hubble time once again and magnetic energy decay can start again~\cite{Banerjee:2004df,Campanelli:2013iaa}. 
The drag force due to the free-streaming of neutrinos becomes ineffective at approximately the freeze-out of neutrinos, i.e. when their mean free path becomes larger than the Hubble scale. The dissipation due to photon drag is efficient until photon freeze-out at \m{T\simeq0.26}~eV when recombination commences.
The decay of magnetic fields due to viscous photon free-streaming is given by 
\m{k_I\propto a^{-3/(n+5)}} and \m{\tilde{\rho}_B\propto a^{-3(n+3)/(n+5)}}~\cite{Banerjee:2004df}. With \m{n=2} we find
\m{\tilde{B}_I\propto a^{-15/14}}, \m{k_I\propto a^{-3/7}} and \m{\tilde{\rho}_B\propto a^{-15/7}},
\begin{comment}
%  
\begin{equation}\label{eq:visc_damp}
 \tilde{B}_I\propto a^{-\frac{15}{14}}\,,\qquad
k_I\propto a^{-\frac37}
\qquad\textrm{and}\qquad
\tilde{\rho}_B\propto a^{-\frac{15}{7}}\,,\qquad\text{(viscous damping, non-helical)}
\end{equation} 
\end{comment}
which is a faster decay than in the turbulent case, and faster than in the case of a maximally helical field. For maximally helical fields the relation \m{\tilde{\rho}_B\propto k_I} again holds, and the decay due to free-streaming photons is
\m{\tilde{B}_I\propto a^{-1/2}}, \m{k_I\propto a^{-1}} and \m{\tilde{\rho}_B\propto a^{-1}}.
\begin{comment}
%  
\begin{equation}\label{eq:visc_damp_max_hel}
 \tilde{B}_I\propto a^{-\frac12}\,,\qquad
k_I\propto a^{-1}
\qquad\textrm{and}\qquad
\tilde{\rho}_B\propto a^{-1}\,,\qquad\text{(viscous damping, max. helical)}
\end{equation}
\end{comment}
We can summarise the above viscous damping (VD) decay laws by
%
\begin{comment}
\begin{equation}\label{eq:visc_decay}
 \tilde{B}_I\propto a^{-\frac32\frac{p+5}{p+7}}\,,\qquad
L_I\propto a^{\frac{3}{p+7}}
\qquad\textrm{and}\qquad
\tilde{\rho}_B\propto a^{-3\frac{p+5}{p+7}}\,,\qquad\text{(viscous damping)}
\end{equation} 
\end{comment}
%
\begin{equation}\label{eq:visc_decay}
(\mathrm{VD}):\qquad
\tilde{\rho}_B\propto a^{-3\frac{p+5}{p+7}}\,,\quad
L_I\propto a^{\frac{3}{p+7}}\,,
\end{equation} 
where \m{p=0} for non-helical fields and \m{p=-4} for maximally helical fields.

In order to calculate the CMB spectral distortions due to decaying magnetic fields, we have to know in which of the above phases of evolution the $\mu$ and $y$ eras correspond to. For this we must understand when these phases begin and end, which is what we do in the next Section.

%---------------------------------------------------
\subsection{End of the turbulent damping stage}\label{sec:EoT_A}

When the magnetic fields are first generated, for example by a first-order phase transition, the magnetic energy is expected to come into equipartition with the kinetic energy [see Ref.~\cite{Wagstaff:2013yna} for a detailed analysis of this mechanism]. At the time of the EW or QCD scale it can be shown that the kinetic Reynolds numbers are very large in this case and the plasma is highly turbulent. Hence magnetogenesis ends with the plasma in a turbulent state. We now calculate \m{T_{\mathrm{EoT}}}, the temperature corresponding to the end-of-turbulence. Turbulence ends when the kinetic Reynolds number decreases to \m{R_e\sim1}, hence the end-of-turbulence temperature is obtained through 
\begin{equation}\label{eq:Re_1}
R_e\left(T_{\mathrm{EoT}}\right)\sim1\,.
\end{equation}

At the time of the EW or QCD scale the plasma viscosity is generated by neutrinos, since at this time they are the particles which are the most efficient at transporting  momentum and heat \cite{Banerjee:2004df}. But as the Universe expands and cools the neutrino mean free path increases which increases the plasma viscosity. Therefore, due to neutrinos, the plasma goes from a turbulent state to a viscous state up until the neutrinos decouple at \m{T_{\nu\text{dec}}\simeq2.6}~MeV. The evolution of magnetic energy and its coherence length in the turbulent stage due to neutrinos followed by the viscous stage before neutrino decoupling can be well approximated by only considering a turbulent damping stage throughout that epoch \cite{Banerjee:2004df}. In our calculations we use this simplifying approximation.

After neutrino decoupling we are in the photon era, when photons generate the plasma shear viscosity $\eta_s$. In this case the Reynolds number is given by \cite{Banerjee:2004df}
\begin{equation}\label{eq:Re}
R_e(l) =\frac{v_l^{\text{rms}}l}{\eta_{s}}
=\frac{5g_*(T)}{g_{\gamma}}
\frac{v_l^{\text{rms}}l_c}{l_{\text{mfp},c}^{\gamma}(T)}\,,
\end{equation} 
for velocity fluctuations $v_l^{\mathrm{rms}}$ correlated on some physical scale $l$. Here $g_*$ and $g_\gamma$ are the total and component number of effective relativistic degrees of freedom. We apply the above from the time of neutrino
decoupling at \m{T_{\nu\text{dec}}\simeq2.6}~MeV to the time of matter-radiation equality at \m{T_{\text{eq}}\simeq1}~eV. In this epoch, the comoving photon mean free path is given by~\cite{Jedamzik:1994dd}
\begin{equation}\label{l_c-photons}
 l_{\text{mfp},c}^\gamma\simeq
\frac{a^{-1}}{\sigma_T\sqrt{n^2_{\textrm{\tiny{pair}}}+n^2_e}}\,,
%\simeq \frac{a^{-1}}{\sigma_T n_{\textrm{\tiny pair}}},
\end{equation}
where \m{\sigma_T=8\pi\alpha^2/3m_e^2} is the
Thomson cross section, \m{\alpha\approx1/137} is
the fine structure constant and $m_e$ is the electron mass.
The number densities 
$n_{\textrm{\tiny pair}}$ and $n_e$ of $e^{\pm}$ pairs and
free electrons respectively are given
by~\cite{Jedamzik:1994dd}
\begin{eqnarray}
 n_{\textrm{\tiny pair}}&\approx&
\left(\frac{2m_e T}{\pi}\right)^{\frac32}\exp\left(-\frac{m_e}{T}\right)
\left(1+\frac{15}{8}\frac{T}{m_e}\right)\,,\label{n_pairs}\\
n_e&=&X_e\frac{\Omega_b\rho_0}{m_\mathrm{pr}}\left(\frac{T}{T_0}
\right)^ { 3 }\,,\label{n_free}
\end{eqnarray}
where
$m_\mathrm{pr}$ is the proton mass, the baryon fraction and present
day density product is
\m{\Omega_b\rho_0\simeq1.81\times10^{-12}~\text{eV}^4}
\cite{Planck_param}, \m{T_0\simeq2.725}~K is the present day
photon temperature and 
the ionization fraction is \m{X_e=1} in the radiation dominated era.
Following the usual assumption that in the turbulent regime \m{R_e(L_I)\gg1} there is equipartition between magnetic and kinetic energy on all scales up to the integral scale, i.e.
\begin{equation}
(v_k^{\mathrm{rms}})^2= \Gamma\frac{(\tilde{B}_k^{\mathrm{rms}})^2}{4\pi\tilde{\rho}}
\qquad\text{for}\qquad k\geq k_I\,.
\end{equation} 
Numerical simulations show that there is almost exact equipartition i.e. \m{\Gamma\approx1} for non-helical fields, which is slightly reduced to \m{\Gamma\approx10^{-1}} for maximally helical fields~\cite{Banerjee:2004df}. Therefore we find on the integral scale
\begin{equation}\label{eq:Re2}
R_e(L_I,T)\simeq
%\sqrt{\Gamma}\frac{5g_*}{g_{\gamma}}
%\frac{L_I}{l_{\text{mfp},c}^{\gamma}}\frac{\tilde{B}_I}{\sqrt{4\pi\tilde{\rho}}}
%=
\sqrt{\Gamma}\frac{5g_*}{g_{\gamma}}
\frac{L_I}{l_{\text{mfp},c}^{\gamma}}
\left(\frac{2\tilde{\rho}_B}{\tilde{\rho}}\right)^{\frac12}\,,
\end{equation} 
which is to be evaluated from the time of neutrino decoupling. To connect the magnetic energy density with the coherence length we use the approximation discussed below Eq.~(\ref{eq:Re_1}). Hence we can use Eq.~(\ref{eq:turb_decay}) to find 
\m{L_I\sqrt{\tilde{\rho}_B}=L_{I,*}\sqrt{\tilde{\rho}_{B,*}}\left(T/T_*\right)^{(p+3)/(p+7)}}, where the index `*' denotes the epoch of magnetogenesis.
%
\begin{comment} 
\begin{equation}
L_I\sqrt{\tilde{\rho}_B}
%=L_{I,*}\sqrt{\tilde{\rho}_{B,*}}\left(\frac{a}{a_*}\right)^{-\frac{p+3}{p+7}}
=L_{I,*}\sqrt{\tilde{\rho}_{B,*}}
\left(\frac{T}{T_*}\right)^{\frac{p+3}{p+7}}
\,.
\end{equation}
\end{comment}
%
With the above the Reynolds number is
\begin{equation}\label{eq:Re_final}
R_e(L_I,T)
%\simeq\sqrt{\Gamma}\frac{5g_*}{g_{\nu,\gamma}}
%\frac{L_{I,*}}{l_{\text{mfp},c}^{\gamma}}
%\left(\frac{2\tilde{\rho}_{B,*}}{\tilde{\rho}}\right)^{\frac12}
%\left(\frac{T}{T_*}\right)^{\frac{p+3}{p+7}}\,,
\simeq\sqrt{2\Gamma\varepsilon}\beta
\frac{5g_*}{g_{\nu,\gamma}}
\frac{\lambda_{B,*}^{\text{max}}}{l_{\text{mfp},c}^{\gamma}}
\left(\frac{T}{T_*}\right)^{\frac{p+3}{p+7}}\,,
\end{equation}
where we have defined \m{\varepsilon\equiv\tilde{\rho}_{B,*}/\rho_{\gamma,0}\simeq\tilde{\rho}_{B,*}/\tilde{\rho}}. Here, the maximal magnetic energy \m{\varepsilon=1} corresponds to \m{u_B=\tilde{\rho}/2}, i.e. a maximum magnetic field strength of
\begin{equation}
 \tilde{B}_{\lambda,*}^{\mathrm{max}}\equiv \sqrt{4\pi\tilde{\rho}}
  \simeq 3\times10^{-6}~\textrm{G}\,,
\end{equation}
where the radiation here is taken to be the CMB photons~\cite{Banerjee:2004df}.
In the above we have also defined  \m{\beta\equiv\lambda_{B,*}/\lambda_{B,*}^{\text{max}}}, and we have identified the integral scale $L_{I,*}$ with the magnetic field coherence length \m{\lambda_{B,*}}.
For magnetic fields generated at a time during the radiation dominated era (in contrast to inflationary magnetogenesis), the basic constraint on the coherence length is the horizon size at the time of magnetogenesis
\begin{equation}
 \lambda_{B,*}\leq \lambda^{\mathrm{max}}_{B,*}\equiv \frac{1}{aH}\Big|_*\,.
\end{equation}
The horizon size is \m{2\times10^{-10}}~Mpc and \m{3\times10^{-7}}~Mpc at the electroweak and QCD  phase transitions respectively. 
To estimate the integral scale at the time of magnetogenesis $L_{I,*}$, we can assume that turbulence is effective such that the Alfv\'{e}n eddy-turnover time $t_{A,*}$ is equal to the Hubble time~\cite{Banerjee:2004df} i.e.
\begin{equation}
 t_{A,*}\equiv\frac{L}{v^{\mathrm{rms}}_{A,L}}\Bigg|_*=\frac{1}{aH}\Big|_*\,,
\end{equation}
where \m{v^{\mathrm{rms}}_{A,L}\equiv\tilde{B}^{\mathrm{rms}}_{L}/\sqrt{4\pi\tilde{\rho}}}. With the above we find that
\m{\lambda_{B,*}=L_{I,*}=\sqrt{\varepsilon}/aH|_*},
hence we can set \m{\beta=\sqrt{\varepsilon}} in Eq.~(\ref{eq:Re_final}) above.

The end-of-turbulence temperature, obtained through 
\m{R_e\left(T_{\mathrm{EoT}}\right)\sim1}, can only be determined numerically; we cannot analytically invert the function in Eq.~(\ref{eq:Re_final}) due to the exponential in the photon mean free path. 
%Reading from the plots in Figure~\ref{fig:Re_T_EoT}, 
We find that \m{T_{\mathrm{EoT}}\simeq2\times10^4}~eV and \m{2\times10^3}~eV for non-helical magnetic fields generated at the EW and QCD phase transition, \m{T_*\simeq100}~GeV and \m{200}~MeV respectively, with \m{\beta=1} and \m{\varepsilon=1}. Hence the $\mu$-era is within the viscous regime, since the $\mu$-era commences at around \m{T_{\mu,i}\simeq470}~eV. However, for maximally helical fields, it is possible that the plasma is still in a turbulent stage within the $\mu$-era. The point here is that the magnetic field decay could be in the turbulent, viscous freezing or viscous damping regime depending on the initial conditions. This can be seen in the evolution plots in Figure~\ref{fig:Mag_evolution_plot}.

\begin{comment}
%
\begin{figure}[ht!]
\includegraphics[width=85mm,trim=0 0 0 0]{Re_T_EoT.pdf}
\includegraphics[width=85mm,trim=0 0 0 0]{Re_T_EoT_helical.pdf}
\caption{\label{fig:Re_T_EoT} Plots (a) and (b) show the temperature as the plasma enters the viscous regime \m{R_e\sim1} for different initial conditions and helicity (non-helical and maximally helical respectively). Since the $\mu$-era commences at around \m{T_{\mu,i}\simeq470}~eV, this plot clearly shows that the $\mu$-era is well within the viscous regime in the non-helical case (a), whereas in the maximally helical case (b) we can see that the system could still be turbulent while entering the $\mu$-era.
}
\end{figure}
%

%
\begin{figure}[ht!]
\includegraphics[width=100mm,trim=0 20 20 20]{Re_T_EoT_helical.pdf}
\caption{\label{fig:Re_T_EoT_helical} Maximally helical case. Reynolds numbers as the plasma enters the viscous regime. Since the $\mu$-era commences at around \m{T\simeq500}~eV, this plot clearly shows that... the $\mu$-era is well within the viscous regime.}
\end{figure}
\end{comment}
%

%---------------------------------------------------
\subsection{Start of the viscous damping stage}

The magnetic energy and coherence length stop evolving in the viscous regime when particles are diffusing \m{l_{\mathrm{mfp},c}\ll L} [see above Eq.~(\ref{eq:visc_freez})]. The viscous damping stage occurs when photons begin to free-stream \m{l_{\mathrm{mfp},c}\gg L}. The start of the viscous damping stage (index `vd') is determined by the condition \cite{Banerjee:2004df}
\begin{equation}\label{eq:tau_visc_free}
 \tau_\mathrm{visc.free}\left(T_\mathrm{vd}\right)=H^{-1}\left(T_\mathrm{vd}\right)\,,
\end{equation}
where the viscous free-streaming time-scale is given by \m{\tau_\mathrm{visc.free}=\alpha_\gamma L^2/v_A^2}. Here, the drag term on the fluid due to the occasional scattering of photons with the fluid particles is given by \cite{Banerjee:2004df}
\begin{equation}
 \alpha_\gamma\simeq\frac43\frac{1}{l_{\mathrm{mfp}}}
\frac{\rho_\gamma}{\rho_b}
\,,
\end{equation}
where $\rho_b$ is the baryon density. The Alfv\'{e}n velocity is given by \m{v_A^2\simeq2\rho_B/\rho_b}, since at this time the photons are decoupled from the fluid. Magnetic dissipation due to photon drag is shown to be efficient until photon decoupling~\cite{Banerjee:2004df}. 

Hence we can solve Eq.~(\ref{eq:tau_visc_free}) for $T_\mathrm{vd}$, the temperature at the start of the viscous damping due to free-streaming photons. We find
\begin{equation}
 T_{\mathrm{vd}}^3\simeq\frac{9}{16\pi^2\alpha^2}
\sqrt{\frac{90}{g_{*,\mathrm{vd}}}}
\frac{m_Pm_{\mathrm{pr}}m_e^2T_0}{X_{e,\mathrm{vd}}\Omega_b\rho_0}
%\left(\lambda_{B,*}^{\mathrm{max}}\right)^{-2}
\left(\frac{T_{\mathrm{EoT}}}{\lambda_{B,*}^{\mathrm{max}} T_*}\right)^2\,,
\end{equation}
which is valid for \m{T\lesssim10^4}~eV when the photon mean free path can be well approximated by \m{l_{\text{mfp}}^\gamma\simeq1/\sigma_T n_e}.

The evolution of magnetic energy due to MHD turbulence occurs approximately until the time of recombination (index `$\mathrm{rec}$'), when 
the field configuration falls on the line given by \cite{Banerjee:2004df,Durrer:2013pga} 
\begin{equation}\label{eq:BLrec}
B_{\lambda,\mathrm{rec}}\simeq8\times10^{-8}
\frac{\lambda_{B,\mathrm{rec}}}{\mathrm{Mpc}}\mathrm{G}\,.
\end{equation}
This line corresponds to the largest eddies being processed at recombination 
\m{1/(aH)|_{\mathrm{rec}}\simeq\lambda/v_A} with $v_A$ the Alfv\'{e}n speed \cite{Jedamzik:1996wp,Banerjee:2004df}. 
Beyond this epoch, the evolution of the field strength and coherence length essentially ceases, with only a logarithmic scaling \cite{Banerjee:2004df}, and the magnetic fields become frozen into the plasma. Hence, fields generated during the radiation era will evolve to fall on the above line at recombination, which are also the values that will be observed today \m{B_{\lambda,\mathrm{rec}}\approx B_0} and \m{\lambda_{B,\mathrm{rec}}\approx\lambda_{B}}, since the field strength and coherence length do not evolve significantly in the matter dominated Universe.

\begin{figure}[ht!]
\includegraphics[width=85mm,trim=1 28 10 0]{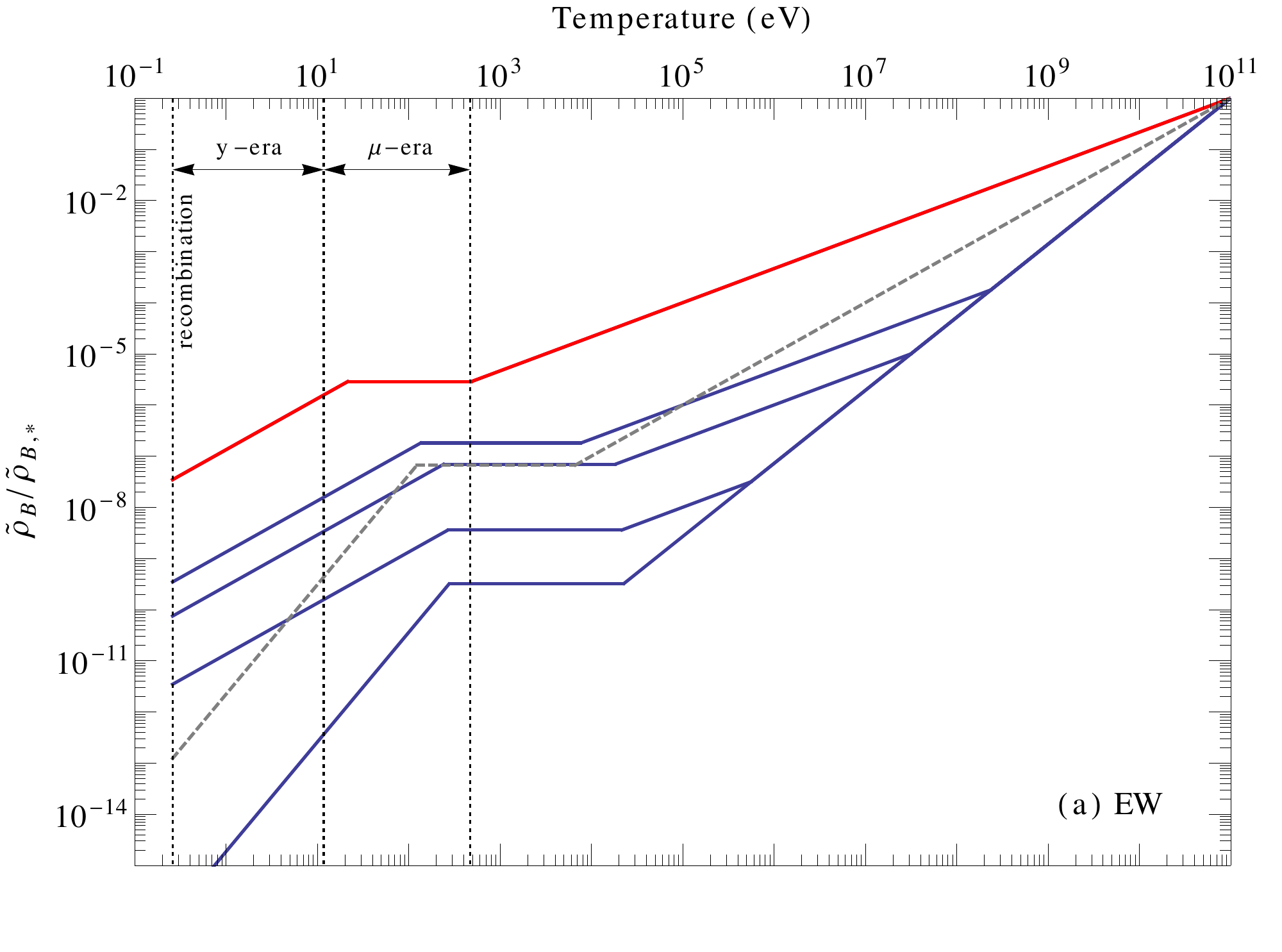}
\includegraphics[width=85mm,trim=0 0 0 28]{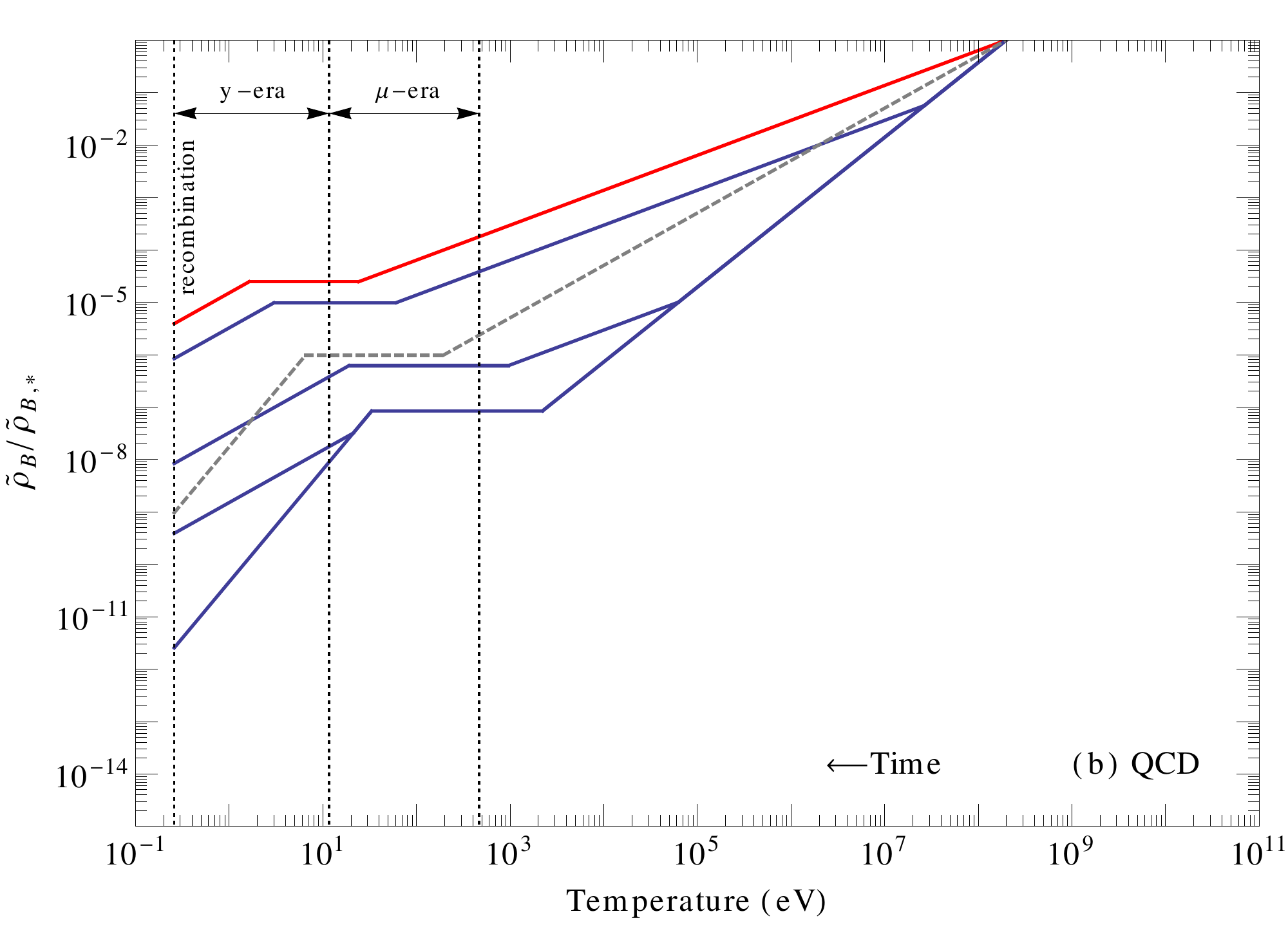}
\caption{\label{fig:Mag_evolution_plot} In plots (a) and (b) we show the evolution of magnetic energy from the time of magnetogenesis (at the EW and QCD scales respectively) to recombination. In plots (a) and (b) the (solid, blue) lines from top to bottom correspond to initial helicity fractions \m{f_*=\left\{10^{-3},10^{-4},10^{-6},<10^{-14}\right\}} and \m{f_*=\left\{10^{-1},10^{-4},10^{-6},<10^{-14}\right\}} respectively.
The maximally helical case \m{f_*=1} (solid, red) is also shown.
We also show (dashed, gray lines) the evolution of non-helical magnetic fields with an inverse transfer of energy (see discussion below Eq.~(\ref{eq:turb_decay})). In all the plots shown we consider \m{\varepsilon\equiv\tilde{\rho}_{B,*}/\rho_{\gamma,0}\approx1}, which corresponds to an initial field strength of \m{\tilde{B}_{\lambda,*}\simeq 3\times10^{-6}~\textrm{G}}, with \m{\varepsilon<1} the evolution history also changes.
}
\end{figure}

We have now all the necessary ingredients to characterise the full evolution of magnetic fields in the radiation dominated epoch. Exemplary evolution histories of magnetic energy for varying initial conditions can be seen in the plots of Figure~\ref{fig:Mag_evolution_plot}, where different initial conditions \m{\tilde{B}_{\lambda,*}}, \m{\lambda_{B,*}}, and \m{f_*} lead to different histories, and hence different decay rates during the $\mu$ and $y$-eras.

%

%------------------------------------------------
%------------------------------------------------
\section{Spectral distortions from decaying magnetic fields}\label{sec:general}

At high temperatures, corresponding to \m{z\gtrsim2\times10^6}, the black-body spectrum of the CMB \cite{CMB_BB} is formed from bremsstrahlung and Double-Compton scattering [see for example Ref.~\cite{KK14} and references therein]. As the redshift drops below \m{2\times10^6} these interactions become inefficient at restoring the black-body spectrum if additional energy is injected into the plasma and distortions could be imprinted from then on. In the early stage \m{2\times10^6 \gtrsim z \gtrsim 5\times10^4} the elastic-Compton scattering is efficient enough and the spectral distortion comes in the form of a non-vanishing chemical potential $\mu$. This $\mu$-type CMB spectral distortion is generated if thermal energy is injected into the plasma during the $\mu$-era defined above. The rate of change of the chemical potential $\mu$ is determined by~\cite{HuSilk:1993,Khatri:2012rt,Chluba:2012gq,Pajer:2013oca}
\begin{equation}\label{eq:dmudt}
\frac{\textrm{d}\mu}{\textrm{d}t}=
-\frac{\mu}{t_{\mathrm{DC}}(z)}+\frac{1.4}{3}\rho_\gamma^{-1}\frac{\textrm{d}Q}{\textrm{d}t}\,,
\end{equation}
where \m{\textrm{d}Q/\textrm{d}t} is the energy injection rate. Here, the time scale for Double-Compton scattering is
\begin{equation}
\frac{t_{\mathrm{DC}}(z)}{\mathrm{s}}=
\frac{2.06\times10^{33}}{\Omega_b h^2}\left(1-\tfrac{1}{2} Y_P\right)^{-1}
%\left(\Omega_b h^2\right)^{-1} 
z^{-\frac92}\,,
\end{equation}
and \m{Y_P=0.24} is the primordial helium mass abundance \cite{Planck2015}. The solution to Eq.~(\ref{eq:dmudt}) is given by~\cite{KK14}
\begin{equation}\label{eq:mu_sol}
\mu=\frac{1.4}{3}\int_{z_i}^{z_\mathrm{end}}
\frac{\textrm{d}z}{\rho_\gamma}\frac{\textrm{d}Q}{\textrm{d}z}
\exp{\left[-\left(\frac{z}{z_{\mathrm{DC}}}\right)^{\frac52}\right]}\,,
\end{equation}
where
\begin{equation}
z_{\mathrm{DC}}\equiv
\frac{1.97\times10^{6}}
{\left(1-\frac12 \frac{Y_P}{0.24}\right)^{\frac52}}
\left(\frac{\Omega_b h^2}{0.0224}\right)^{-\frac25}\,,
\end{equation}
and where \m{z_i=2\times10^6} and \m{z_\mathrm{end}=5\times10^4} define the start and end of the $\mu$-era.

%----------------------------------------------------
%----------------------------------------------------

\subsection{CMB $\mu$-type distortions from decaying magnetic fields}

From the full evolution history of magnetic fields shown in Figure \ref{fig:Mag_evolution_plot} we can see that, in most cases, the plasma is in the viscous regime during the $\mu$-era. This is true in most cases except for fields generated at the QCD scale if the initial helicity fraction is large enough. For analytical purposes let us first consider the viscous damping law, from which we can write
\begin{equation}
\tilde{\rho}_B(z) 
%= \tilde{\rho}_{B,\mathrm{vd}}\left(\frac{a}{a_{\mathrm{vd}}}\right)^{-3\frac{p+5}{p+7}}
%=\tilde{\rho}_{B,\mathrm{vd}}\left(\frac{T}{T_{\mathrm{vd}}}\right)^{3\frac{p+5}{p+7}}
=\tilde{\rho}_{B,\mathrm{vd}}
\left(\frac{1+z}{1+z_{\mathrm{vd}}}\right)^{3\frac{p+5}{p+7}}\,,
\end{equation}
%
\begin{comment}
\begin{equation}
\tilde{\rho}_B(z) = \tilde{\rho}_{B,\mathrm{vd}}\left(\frac{a}{a_{\mathrm{vd}}}\right)^{-\frac{15}{7}}
=\tilde{\rho}_{B,\mathrm{vd}}\left(\frac{T}{T_{\mathrm{vd}}}\right)^{\frac{15}{7}}
=\tilde{\rho}_{B,\mathrm{vd}}\left(\frac{1+z}{1+z_{\mathrm{vd}}}\right)^{\frac{15}{7}}\,,
\end{equation}
\end{comment}
where \m{\tilde{\rho}_{B,\mathrm{vd}}} indicates the magnetic energy at the start of the viscous damping stage. Here we also use \m{a\propto1/T} for the photon temperature and \m{T=T_0\left(1+z\right)}. Since the magnetic energy is frozen-out from the time of the end-of-turbulence (EoT) to the start of the viscous damping stage (vd) we can set \m{\tilde{\rho}_{B,\mathrm{vd}}\simeq\tilde{\rho}_{B,\mathrm{EoT}}}, hence we find
\begin{eqnarray}\label{eq:dQdz2}
\rho_{\gamma}^{-1}\frac{\textrm{d}Q}{\textrm{d}z}&=&
-\rho_{\gamma,0}^{-1}\frac{\textrm{d}\tilde{\rho}_B}{\textrm{d}z}\\
&\simeq&-3\frac{p+5}{p+7}
\frac{\tilde{\rho}_{B,\mathrm{EoT}}}{\rho_{\gamma,0}}
\left(1+z_{\mathrm{vd}}\right)^{-3\frac{p+5}{p+7}}
\left(1+z\right)^{2\frac{p+4}{p+7}}\,.\nonumber
\end{eqnarray}
%
\begin{comment}
\begin{equation}\label{eq:dQdz2a}
\rho_{\gamma}^{-1}\frac{\textrm{d}Q}{\textrm{d}z}=
-\rho_{\gamma,0}^{-1}\frac{\textrm{d}\tilde{\rho}_B}{\textrm{d}z}
\simeq-\frac{15}{7}\frac{\tilde{\rho}_{B,\mathrm{EoT}}}{\rho_{\gamma,0}}
\left(1+z_{\mathrm{vd}}\right)^{-\frac{15}{7}}
\left(1+z\right)^{\frac{8}{7}}\,.
\end{equation}
\end{comment}
%
\begin{comment}
Therefore the $\mu$-distortion in Eq.~(\ref{eq:mu_sol}) is given by
%
\begin{equation}\label{eq:mu_sol2}
\mu=-\frac{\epsilon_{B,\mathrm{EoT}}}{\rho_{\gamma,0}}
\left(1+z_{\mathrm{vd}}\right)^{-\frac{30}{14}}\int_{z_1}^{z_2}\textrm{d}z
\left(1+z\right)^{\frac{16}{14}}
\exp{\left[-\left(\frac{z}{z_{\mathrm{DC}}}\right)^{\frac52}\right]}\,.
\end{equation}
\end{comment}
%
To calculate \m{\tilde{\rho}_{B,\mathrm{EoT}}} we consider the era when turbulence is generated by photons after neutrino decoupling. As already argued in the Section \ref{sec:EoT_A}, we can trace the evolution of the magnetic energy all the way back to the time of magnetogenesis using the turbulent decay law given in Eq.~(\ref{eq:turb_decay}). This is possible since the turbulent stage due to neutrinos followed by the viscous stage before neutrino decoupling can be well approximated by only considering a turbulent damping stage throughout the epoch \cite{Banerjee:2004df}. Hence, using the turbulent damping decay law, we can write
\begin{equation}\label{eq:rho_EoT}
\tilde{\rho}_{B,\mathrm{EoT}}
%=\tilde{\rho}_{B,*}\left(\frac{a_{\mathrm{EoT}}}{a_{*}}\right)^{-2\frac{p+5}{p+7}}
=\tilde{\rho}_{B,*}\left(\frac{T_{\mathrm{EoT}}}{T_{*}}\right)^{2\frac{p+5}{p+7}}\,.
\end{equation}
%
\begin{comment}
\begin{equation}\label{eq:rho_EoTa}
\tilde{\rho}_{B,\mathrm{EoT}}
=\tilde{\rho}_{B,*}\left(\frac{a_{\mathrm{EoT}}}{a_{*}}\right)^{-\frac{10}{7}}
=\tilde{\rho}_{B,*}\left(\frac{T_{\mathrm{EoT}}}{T_{*}}\right)^{\frac{10}{7}}\,.
\end{equation}
\end{comment}
%
In the cases where the viscous regime starts before the $\mu$-era, i.e. \m{T_{\mathrm{EoT}}\geq T_{\mu,i}}, we can use Eqs.~(\ref{eq:dQdz2}) and (\ref{eq:rho_EoT}) to estimate the $\mu$-type distortion from Eq.~(\ref{eq:mu_sol}), we find
\begin{eqnarray}\label{eq:mu_sol3}
\mu=-&&\frac{7}{5}
\frac{p+5}{p+7}
\left(\frac{\tilde{\rho}_{B,*}}{\rho_{\gamma,0}}\right)
\left(\frac{T_{\mathrm{EoT}}}{T_{*}}\right)^{2\frac{p+5}{p+7}}
\left(1+z_{\mathrm{vd}}\right)^{-3\frac{p+5}{p+7}}\times\nonumber\\
&&\times\int_{z_i}^{z_\mathrm{end}}\textrm{d}z
\left(1+z\right)^{2\frac{p+4}{p+7}}
\exp{\left[-\left(\frac{z}{z_{\mathrm{DC}}}\right)^{\frac52}\right]}\,,
\end{eqnarray}
where we integrate from \m{z_i=z_{\mathrm{vd}}} to $z_\mathrm{end}$. 
In the cases where the turbulent regime has not ended by the start of the $\mu$-era, e.g. applicable for fully helical magnetic fields generated at the QCD scale with \m{\varepsilon\simeq1-10^{-2}} as can be seen in figure~\ref{fig:Mag_evolution_plot}, we find
\begin{eqnarray}\label{eq:mu_sol4}
\mu=-&&\frac{14}{15}
\frac{p+5}{p+7}
\left(\frac{\tilde{\rho}_{B,*}}{\rho_{\gamma,0}}\right)
\left(\frac{T_0}{T_*}\right)^{2\frac{p+5}{p+7}}\times\\
&&\times\int_{z_i}^{z_\mathrm{end}}\textrm{d}z
\left(1+z\right)^{\frac{p+3}{p+7}}
\exp{\left[-\left(\frac{z}{z_{\mathrm{DC}}}\right)^{\frac52}\right]}\,,\nonumber
\end{eqnarray}
where we integrate from \m{z_i} to \m{z_\mathrm{end}=z_\mathrm{EoT}}.

\begin{comment}
\begin{equation}\label{eq:mu_sol3a}
\mu=-\frac{\tilde{\rho}_{B,*}}{\rho_{\gamma,0}}
\left(\frac{T_{\mathrm{EoT}}}{T_{*}}\right)^{\frac{10}{7}}
\left(1+z_{\mathrm{vd}}\right)^{-\frac{15}{7}}
\int_{z_i}^{z_\mathrm{end}}\textrm{d}z
\left(1+z\right)^{\frac{8}{7}}
\exp{\left[-\left(\frac{z}{z_{\mathrm{DC}}}\right)^{\frac52}\right]}\,.
\end{equation}
\end{comment}

\begin{comment}
In the maximally helical case \m{p=-4}, the spectral distortion is
%
\begin{equation}\label{eq:mu_sol3b}
\mu=-\frac{7}{15}
\left(\frac{\tilde{\rho}_{B,*}}{\rho_{\gamma,0}}\right)
\left(\frac{T_{\mathrm{EoT}}}{T_{*}}\right)^{\frac{2}{3}}
\left(1+z_{\mathrm{vd}}\right)^{-1}
\int_{z_i}^{z_\mathrm{end}}\textrm{d}z
\exp{\left[-\left(\frac{z}{z_{\mathrm{DC}}}\right)^{\frac52}\right]}\,.
\end{equation}
\end{comment}

\begin{figure}[ht!]
\includegraphics[width=85mm,trim=0 29 8 0]{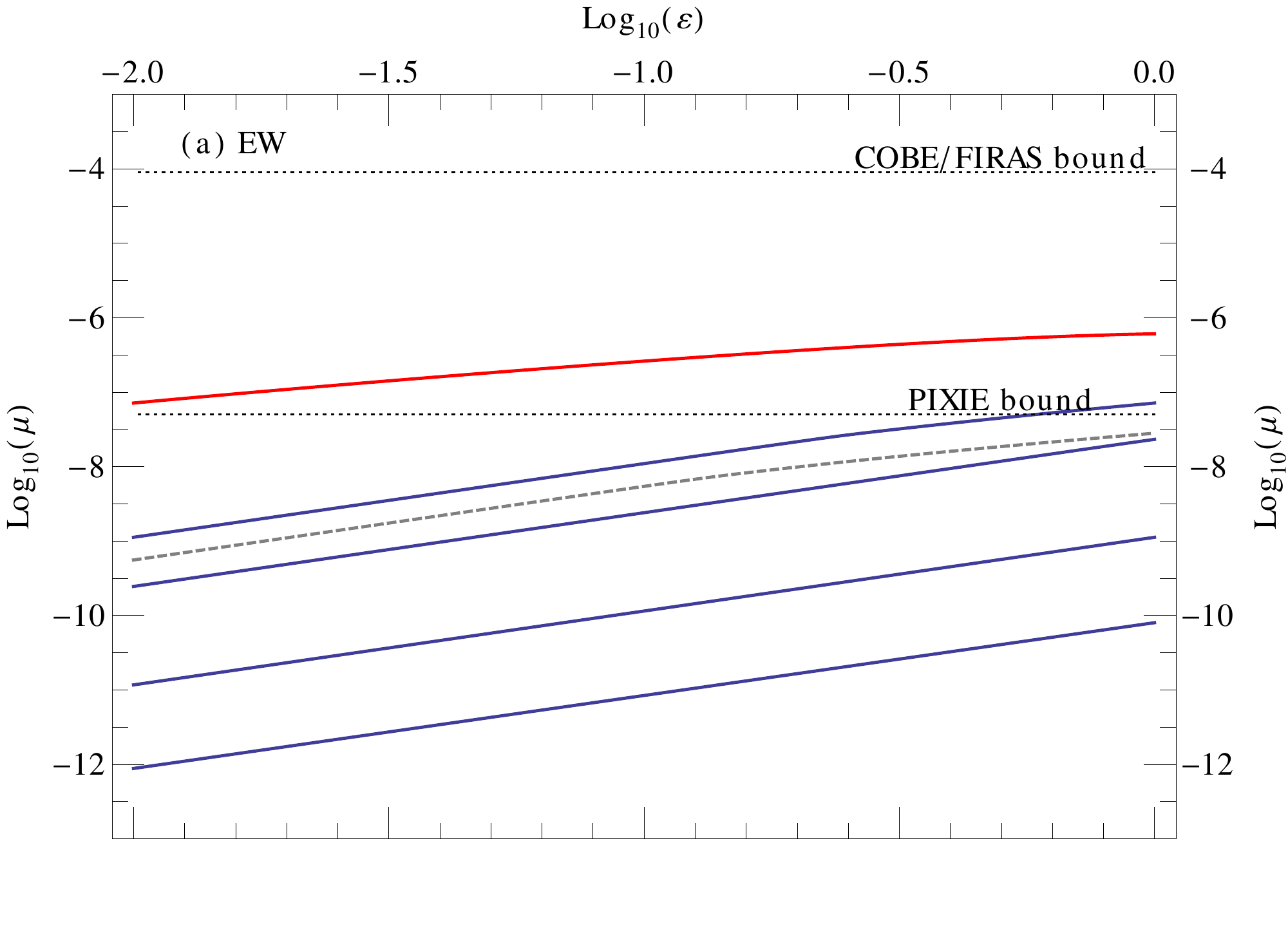}
\includegraphics[width=85mm,trim=0 0 0 29]{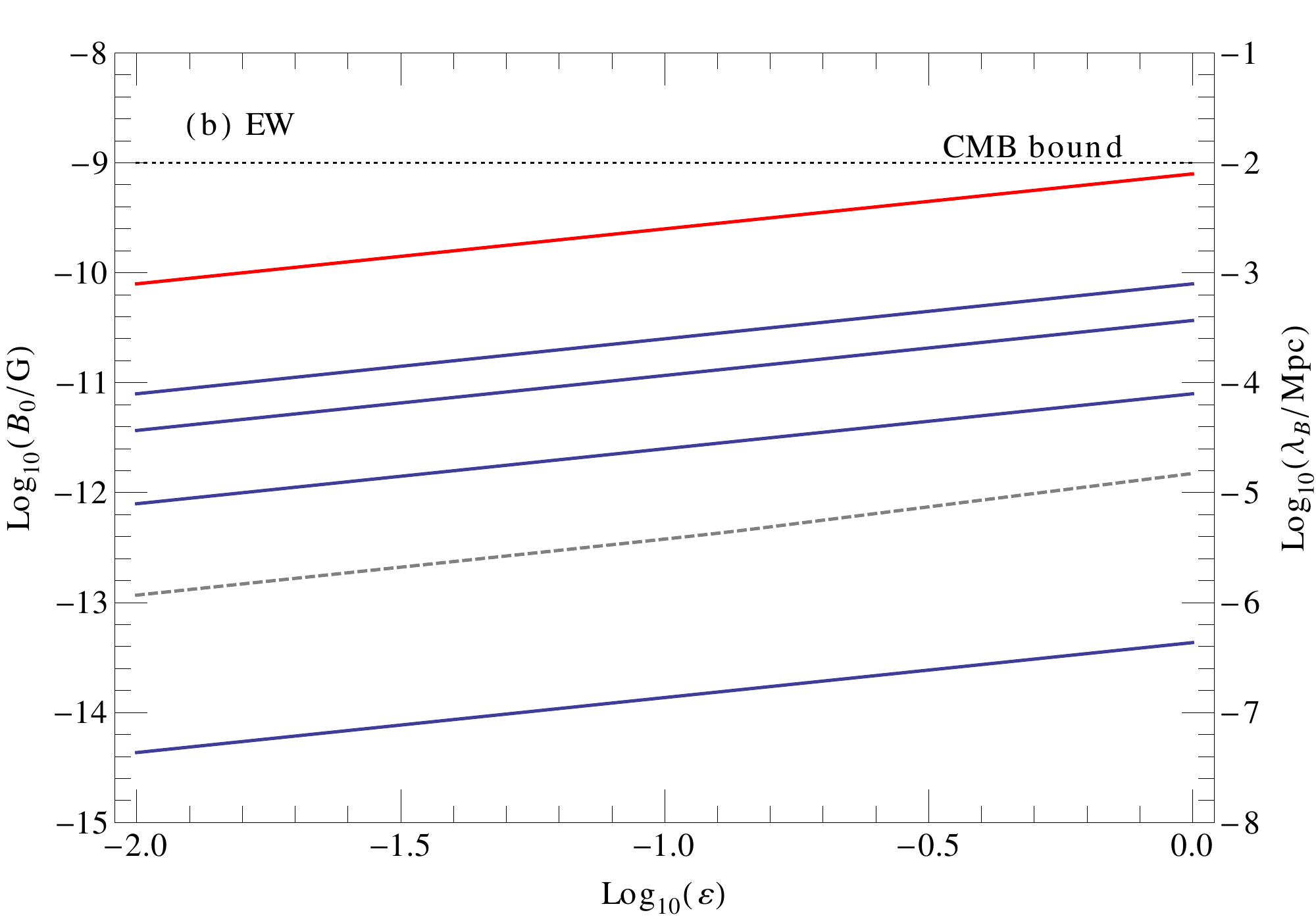}
\caption{\label{fig:mu_z_plot_EW}
EW scale: In plot (a) we show the $\mu$-type distortion generated due to the decay of magnetic energy initially generated at the EW scale. Here we plot the spectral distortion $\mu$ vs $\varepsilon$, where \m{\varepsilon\equiv\tilde{\rho}_{B,*}/\rho_{\gamma,0}\approx1} corresponds to an initial field strength \m{\tilde{B}_{\lambda,*}\simeq 3\times10^{-6}~\textrm{G}}.
The (solid, blue) lines from top to bottom, in both plots, correspond to initial helicity fractions \m{f_*=\left\{10^{-3},10^{-4},10^{-6},<10^{-14}\right\}}. 
The maximally helical case \m{f_*=1} (solid, red) is also shown.
In plot (b) we show the final field strength $B_0$ and coherence length $\lambda_B$ that would be observed today, i.e. after MHD turbulent decay, see Eq.~(\ref{eq:BLrec}). 
We also show the approximate constraint on magnetic fields from CMB observations, \m{B_0\lesssim10^{-9}}G see Ref.~\cite{Planck2015Mag} and references therein. The results for non-helical magnetic fields with an inverse transfer of energy (see discussion below Eq.~(\ref{eq:turb_decay})) are also shown (dashed, gray lines).
}
\end{figure}

\begin{figure}[ht!]
\includegraphics[width=85mm,trim=0 29 8 0]{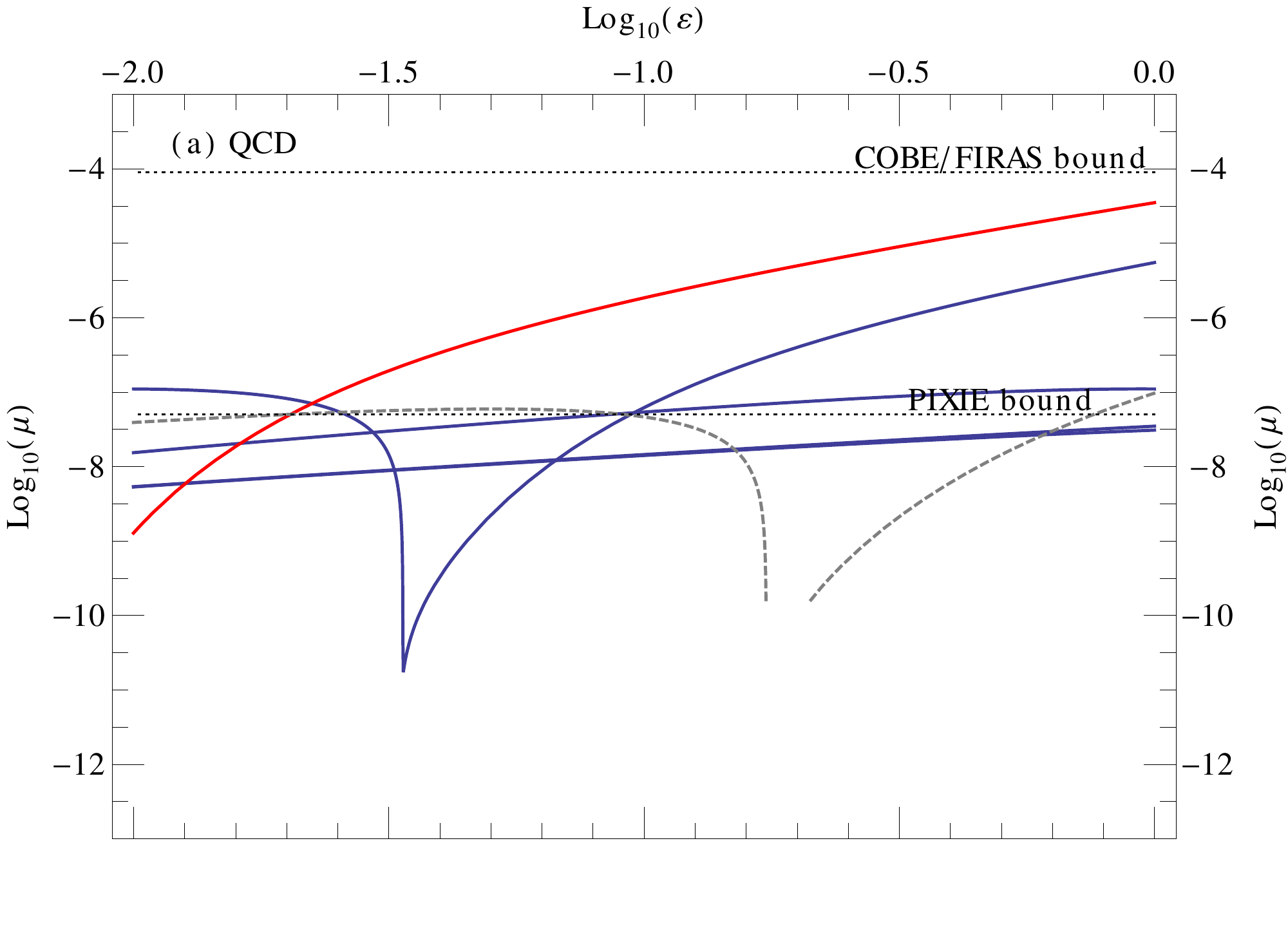}
\includegraphics[width=85mm,trim=0 0 0 29]{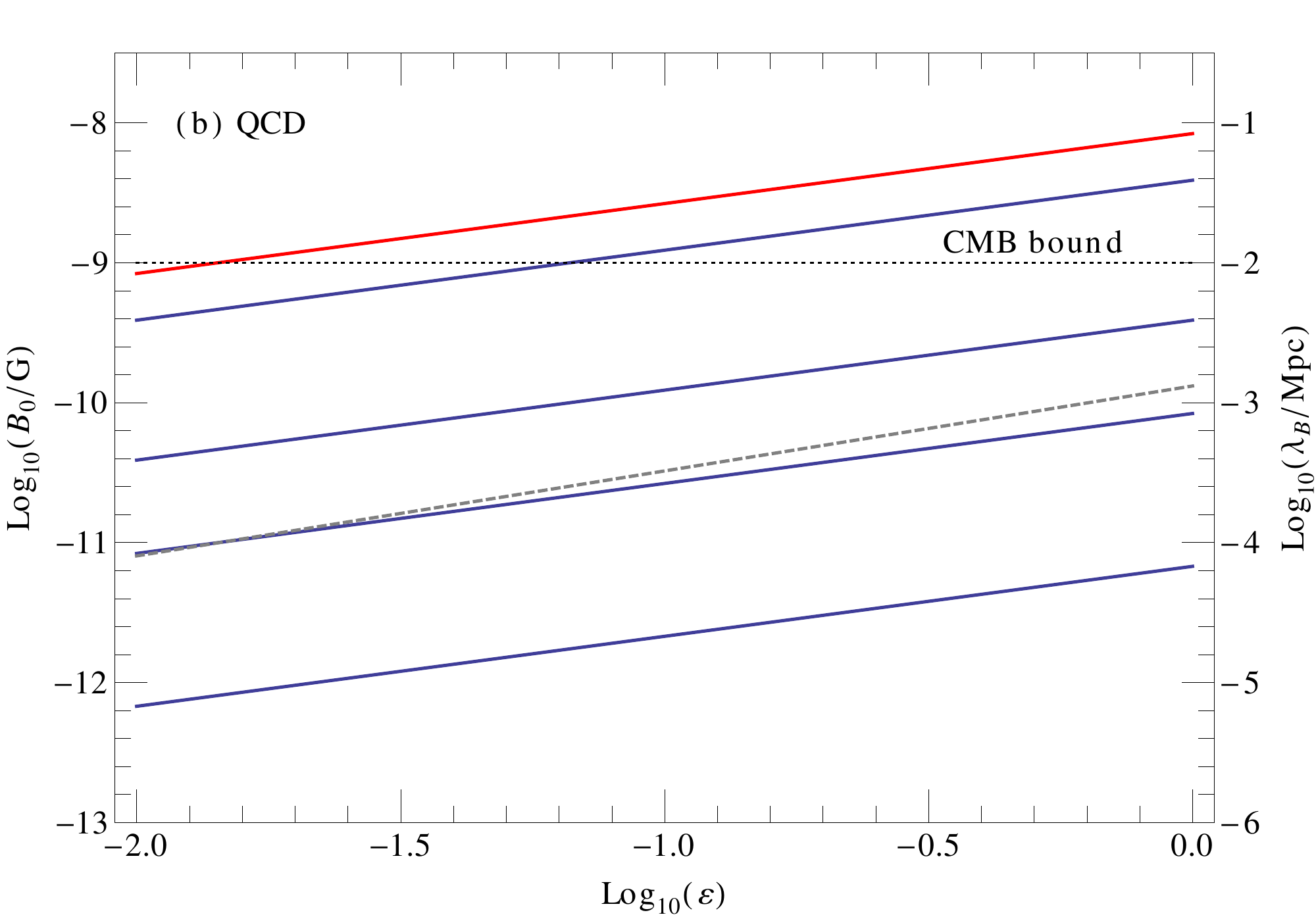}
\caption{\label{fig:mu_z_plot_QCD} 
QCD scale: In plot (a) we show the $\mu$-type distortion generated due to the decay of magnetic energy initially generated at the QCD scale. Here we plot the spectral distortion $\mu$ vs $\varepsilon$, where \m{\varepsilon\equiv\tilde{\rho}_{B,*}/\rho_{\gamma,0}\approx1} corresponds to an initial field strength \m{\tilde{B}_{\lambda,*}\simeq 3\times10^{-6}~\textrm{G}}.
The (solid, blue) lines from top to bottom, in both plots, correspond to initial helicity fractions \m{f_*=\left\{10^{-1},10^{-4},10^{-6},<10^{-14}\right\}}. The maximally helical case \m{f_*=1} (solid, red) is also shown.
In plot (b) we show the final field strength $B_0$ and coherence length $\lambda_B$ that would be observed today, i.e. after MHD turbulent decay, see Eq.~(\ref{eq:BLrec}). We also show the approximate constraint on magnetic fields from CMB observations, \m{B_0\lesssim10^{-9}}G see Ref.~\cite{Planck2015Mag} and references therein. The results for non-helical magnetic fields with an inverse transfer of energy (see discussion below Eq.~(\ref{eq:turb_decay})) are also shown (dashed, gray lines). Here we comment on the seemingly strange behaviour of the plot for \m{f_*=10^{-1}} in the QCD case (also applicable to the non-helical inverse transfer case, gray dashed line). As $\varepsilon$ decreases the onset of the viscous freezing regime begins earlier. For a certain value of $\varepsilon$ the start of the viscous freezing regime coincides with the start of the $\mu$-era, hence there is practically no magnetic energy decay in this time and hence $\mu$ becomes very small. As $\varepsilon$ decrease further, the start of the viscous damping stage occurs within the $\mu$-era and a large $\mu$ can once again be generated.
}
\end{figure}

Equations (\ref{eq:mu_sol3}) and (\ref{eq:mu_sol4}) above are valid for either non-helical fields (\m{p=0}) or maximally-helical fields (\m{p=-4}). Since we are interested in different initial conditions, in particular varying initial helicity fractions $f_*$, we must consider the full evolution history to calculate the spectral distortions. This is done numerically and the results are shown in Figures~\ref{fig:mu_z_plot_EW} and \ref{fig:mu_z_plot_QCD} for magnetic fields generated at the EW and QCD scales respectively. The $\mu$-type spectral distortion varies with \m{\mu=\mu\left(\varepsilon,T_*,f_*\right)}. In order to maximise this distortion we can set the maximal value \m{\varepsilon\equiv\tilde{\rho}_{B,*}/\rho_{\gamma,0}\approx1} at the time of magnetogenesis, which corresponds to an initial field strength of \m{\tilde{B}_{\lambda,*}\simeq 3\times10^{-6}~\textrm{G}}. From Eq.~(\ref{eq:mu_sol3}) we see that if \m{\varepsilon<1} then \m{T_{\mathrm{EoT}}} will be larger, thereby increasing the chemical potential. However, \m{T_{\mathrm{EoT}}} does not depend so strongly on $\varepsilon$, therefore $\varepsilon$ should be maximised in order to maximise $\mu$. Hence, with \m{\varepsilon=1}, integrating Eq.~(\ref{eq:mu_sol3}) we obtain upper limits from Figures~\ref{fig:mu_z_plot_EW} and \ref{fig:mu_z_plot_QCD}. For fields generated at the EW phase transition \m{T_*\simeq100}~GeV, see Figure~\ref{fig:mu_z_plot_EW}, with \m{f_*\lesssim10^{-14}} we find
\begin{equation}
 |\mu|\lesssim 1\times10^{-10}\,.
\end{equation}
The above satisfies the current COBE/FIRAS limit \m{|\mu|<9\times10^{-5}} \cite{CMB_BB}, but will also not be detectable by a new PIXIE-like experiment which would place a new upper limit of \m{|\mu|<5\times10^{-8}} \cite{PIXIE}. From this we conclude that causally generated non-helical magnetic fields at the electroweak phase transition will not produce any detectable CMB $\mu$-type spectral distortions. 
This is true even if we consider the non-helical inverse transfer effect seen in Refs.~\cite{Kahniashvili:2012uj,Brandenburg:2014mwa} and discussed below Eq.~(\ref{eq:turb_decay}), see the gray dashed line in Figure \ref{fig:mu_z_plot_EW}.
For non-helical fields generated at the QCD phase transition \m{T_*\simeq200}~MeV, see Figure \ref{fig:mu_z_plot_QCD}, we obtain the upper limit \m{|\mu|\lesssim 3\times10^{-8}}, which satisfies the current COBE/FIRAS limit \cite{CMB_BB}, and is very much on the limit of detectability by a new PIXIE-like experiment \cite{PIXIE}.
The results do not change much if we consider the non-helical inverse transfer effect discussed below Eq.~(\ref{eq:turb_decay}), see the gray dashed line in Figure \ref{fig:mu_z_plot_QCD}.


The situation is quite different if magnetic helicity is introduced. As can be seen from Figures \ref{fig:mu_z_plot_EW} and \ref{fig:mu_z_plot_QCD}, if the initial helicity fraction \m{f_*\gtrsim\left(10^{-3}-10^{-4}\right)} then PIXIE-observable $\mu$-type distortions can be generated. We also note that for maximally-helical fields \m{f_*=1} the current COBE/FIRAS limit is not broken \cite{CMB_BB}, so that we cannot constrain primordial helicity from current data. If a future PIXIE-like experiment positively detects a $\mu$-type spectral distortion then primordial magnetic helicity can be constrained.

%%----------------------------------------------------
Here we note that in Refs.~\cite{KK14,Jedamzik_CMBdist} authors considered the evolution of the photon diffusion scale, i.e. the Silk damping scale, as the evolution of the damping scale $k_d$. Where we can write \m{k_I\approx k_d} with an equivalent evolution. However in these works, the authors only considered a scale-invariant Alfv\'{e}n velocity, corresponding to scale-invariant spectrum \m{n=-3}, this gives the much faster evolution of the integral scale \m{k_I\propto a^{-3/2}} (for non-helical fields with \m{p=0}) as seen in their paper~\cite{KK14}. However, the magnetic energy is still considered to evolve along the large scale spectrum, i.e. \m{\tilde{B}_I\propto k_I^{5/2}}, hence the authors find a much faster magnetic field decay rate \m{\tilde{\rho}_B\propto a^{-15/2}}. With this decay rate, the chemical potential scales as \m{\mu\propto \int\textrm{d}z(1+z)^{13/2}
e^{-(z/z_{\mathrm{DC}})^{5/2}}} as seen in their paper~\cite{KK14}. This fast magnetic field decay rate leads to a huge overestimation of the generated chemical potential for the causally generated magnetic fields. In our study, as can be seen in Figures \ref{fig:mu_z_plot_EW} and \ref{fig:mu_z_plot_QCD}, the maximum possible magnetic energy \m{\varepsilon=1}, which corresponds to \m{\tilde{B}_{\lambda,*}\simeq3\times10^{-6}~\textrm{G}} at magnetogenesis, does not over-generate spectral distortions. However, in the work of Ref.~\cite{KK14}, an upper limit of \m{\sim10^{-11}}~nG on the field strength of non-helical magnetic fields is obtained due to current spectral distortions constraints.

%%---------------------------------------------------------
%%---------------------------------------------------------

\subsection{CMB $y$-type distortions from decaying magnetic fields}

After the end of the $\mu$-era, for \m{z\lesssim 5\times10^4}, the elastic Compton scattering becomes ineffective. From then on the spectral distortion becomes a $y$-type i.e. defined by the Compton $y$-parameter~\cite{Chluba:2012gq}
\begin{equation}\label{eq:y_sol}
y=\frac{1}{12}\int^{z_\mathrm{dec}}_{z_\mathrm{end}}
\frac{\textrm{d}z}{\rho_\gamma}\frac{\textrm{d}Q}{\textrm{d}z}\,,
\end{equation}
where \m{z_\mathrm{eq}\simeq3265} and \m{z_\mathrm{dec}\simeq1088} for the time of matter-radiation equality and decoupling respectively. It is possible that intermediate distortions between $\mu$ and $y$-type are generated \cite{Chluba:2011hw,Khatri:2012tw}, but for the purpose of this paper we will only calculate the $\mu$ and $y$ defined above.

\begin{figure}[ht!]
\includegraphics[width=85mm,trim=2 28 19 0]{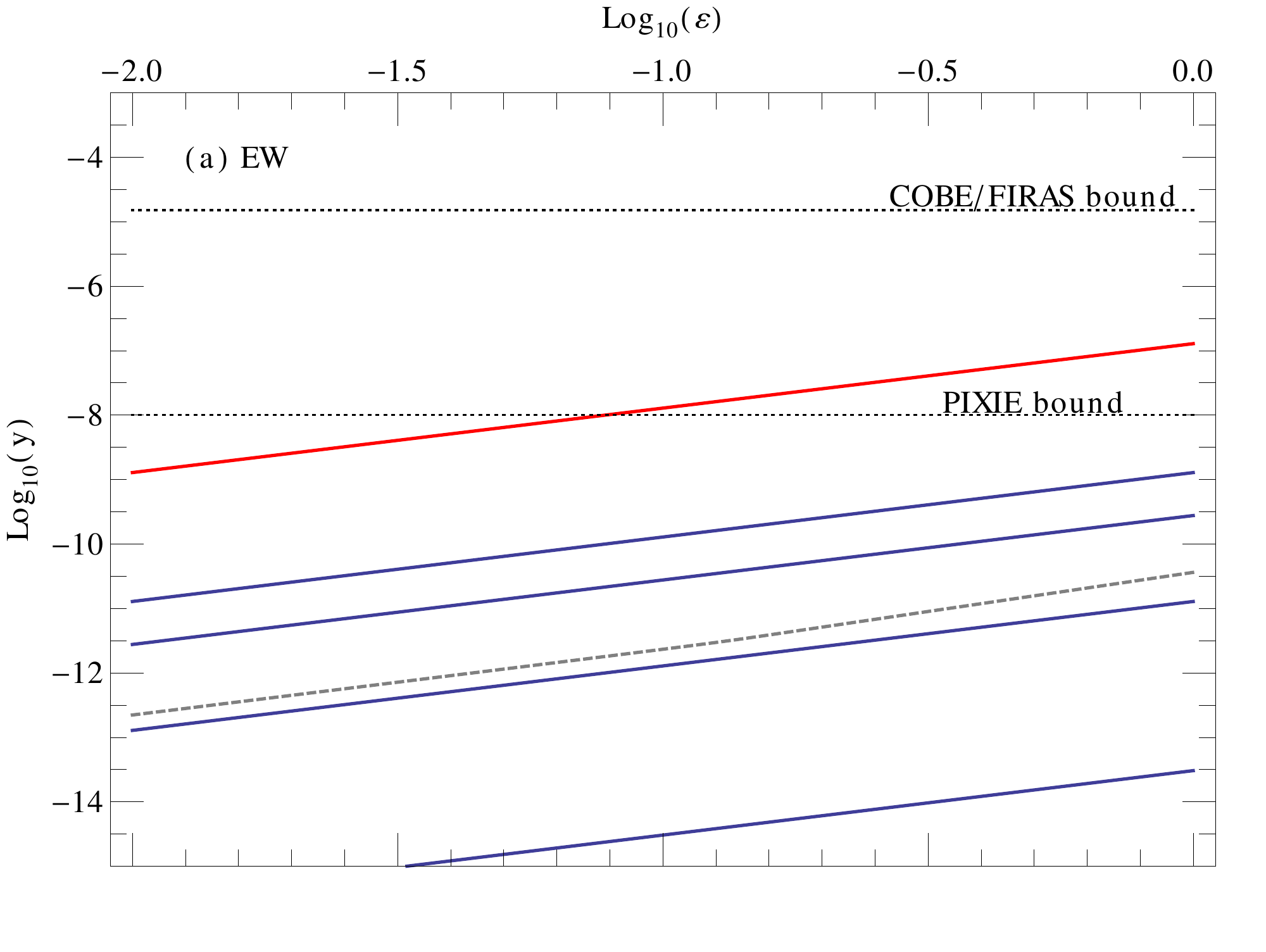}
\includegraphics[width=85mm,trim=0 0 0 28]{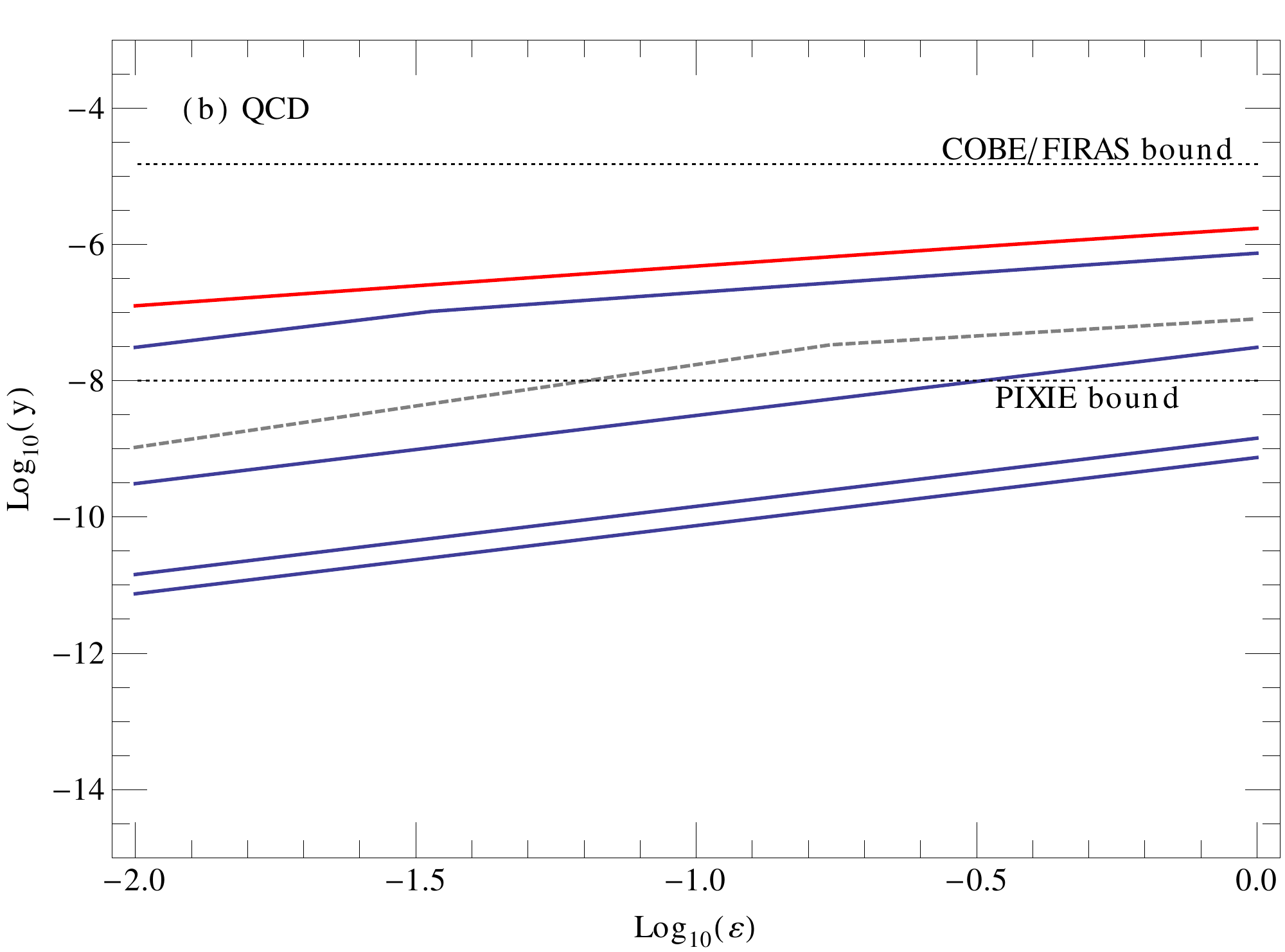}
\caption{\label{fig:y_z_plot} 
In plots (a) and (b) we show the $y$-type distortion produced due to the decay of magnetic energy initially generated at the EW and QCD scales respectively. 
Here we plot the spectral distortion $y$ vs $\varepsilon$, where \m{\varepsilon\equiv\tilde{\rho}_{B,*}/\rho_{\gamma,0}\approx1} corresponds to an initial field strength of \m{\tilde{B}_{\lambda,*}\simeq 3\times10^{-6}~\textrm{G}}.
In plots (a) and (b) the (solid, blue) lines from top to bottom correspond to initial helicity fractions \m{f_*=\left\{10^{-3},10^{-4},10^{-6},<10^{-14}\right\}} and \m{f_*=\left\{10^{-1},10^{-4},10^{-6},<10^{-14}\right\}} respectively. The maximally helical case \m{f_*=1} (solid, red) is also shown. The final field strength $B_0$ and coherence length $\lambda_B$ that would be observed today, i.e. after MHD turbulent decay, see Eq.~(\ref{eq:BLrec}), are the same as in plots (b) of Figures \ref{fig:mu_z_plot_EW} and \ref{fig:mu_z_plot_QCD}. 
We also show (dashed, gray lines) the results from non-helical magnetic fields with an inverse transfer of energy (see discussion below Eq.~(\ref{eq:turb_decay})).
}
\end{figure}

We can see from the plots in Figure \ref{fig:Mag_evolution_plot} that in all cases, for magnetic fields generated at the EW or QCD scale and for any initial helicity fraction, the plasma is in a viscous state (viscous freezing or viscous damping) in the $y$-era. Hence, as a first approximation for the maximum $y$-type distortion generated we can assume the viscous damping law given in Eq.~(\ref{eq:visc_decay}) and a radiation dominated universe throughout the $y$-era. From Eq.~(\ref{eq:y_sol}) and the above considerations we find
\begin{eqnarray}\label{eq:y_sol3}
y=-\frac{1}{4}&&\frac{p+5}{p+7}
\left(\frac{\tilde{\rho}_{B,*}}{\rho_{\gamma,0}}\right)
\left(\frac{T_{\mathrm{EoT}}}{T_{*}}\right)^{2\frac{p+5}{p+7}}\times\\
&&\times\left(1+z_{\mathrm{vd}}\right)^{-3\frac{p+5}{p+7}}
\int^{z_\mathrm{dec}}_{z_\mathrm{end}}\textrm{d}z
\left(1+z\right)^{2\frac{p+4}{p+7}}
\,.\nonumber
\end{eqnarray}
Again, the above expression is valid for non-helical fields (\m{p=0}) or maximally helical fields (\m{p=-4}). For varying initial helicity fractions $f_*$ a full calculation is done taking into acound the full evolution history. The results are shown in the plots of Figure \ref{fig:y_z_plot}.

For non-helical fields, we can see from Figure~\ref{fig:y_z_plot} that the maximum possible $y$-type distortion generated, corresponding to fields generated at the QCD scale, is \m{y\lesssim 8\times10^{-10}}. This result satisfies the current COBE/FIRAS limit \m{y\lesssim1.5\times10^{-5}} \cite{CMB_BB}, and is probably not detectable by a new PIXIE-like experiment, which would place a new lower bound at \m{y\lesssim10^{-8}} if not detected \cite{PIXIE}. For non-helical fields generated at the EW phase transition there is little hope for detection with the maximum distortion at \m{y\simeq 3\times10^{-14}}. We note that, if we consider the non-helical inverse transfer effect discussed below Eq.~(\ref{eq:turb_decay}) [see the gray dashed line in Figure \ref{fig:y_z_plot}], the $y$-type distortion is still undetectable for fields generated at the EW scale, however it seems now possible to detect a $y$-type distortion if the fields are generated at the QCD scale.

For decaying helical magnetic fields there is a possibility to detect the $y$-type distortion. However, since the $y$-era is later in the evolution history, magnetic fields have already substantially decayed and thus generate a smaller signal than that for the $\mu$-type distortion. Hence $y$-type distortions will always provide weaker constraints than $\mu$-type distortions.

%

%%---------------------------------------------------------
%%---------------------------------------------------------

\section{Conclusions}

Magnetic fields generated in the very early Universe decay in the radiation dominated epoch due to turbulent MHD effects. The decaying magnetic fields inject energy into the primordial plasma which can lead to $\mu$-type and $y$-type distortions to the CMB black body spectrum. The current COBE/FIRAS limits on these spectral distortions are very tight \m{|\mu|<9\times10^{-5}} and \m{y\lesssim1.5\times10^{-5}} \cite{CMB_BB}. However there is the exciting possibility of a new PIXIE-like experiment which could place much stronger upper limits of \m{|\mu|<5\times10^{-8}} and \m{y\lesssim10^{-8}} if no detection is made \cite{PIXIE}. Any prediction for spectral distortions above the PIXIE limits is what we call detectable.

In this paper we consider the evolution of helical and non-helical magnetic fields generated by some causal process in the early Universe. We calculate the spectral distortions using the decays laws of Refs.~\cite{Banerjee:2004df,Campanelli:2007tc,Campanelli:2013iaa}. 
We find that causally generated non-helical magnetic fields, with an initial helicity fraction less than \m{\sim10^{-14}}, generated at the EW phase transition will not produce any detectable CMB $\mu$-type or $y$-type spectral distortions. 
This remains true even if the inverse transfer effect for non-helical fields seen in Refs.~\cite{Kahniashvili:2012uj,Brandenburg:2014mwa} is considered. Hence, to produce observable spectral distortions from the decay of magnetic fields generated at the EW phase transition, a non-negligible helical component is required.

Here we note that, if the inverse transfer effect for non-helical fields is applicable \cite{Kahniashvili:2012uj,Brandenburg:2014mwa}, it looks possible to generate small amounts of detectable distortions from magnetic fields generated at the QCD phase transition. 
%However we note two points on the inverse transfer effect for non-helical fields; first the effect may not be efficient in the early Universe where the Prandtl numbers are huge, and secondly the effect may only be applicable to magnetically dominant turbulence which is probably not satisfied for fields generated at first order phase transitions where the turbulent kinetic energy is very large. 
We also note that magnetogenesis at the QCD phase transition is disfavoured compared to magnetogenesis at the EW phase transition. Under early Universe conditions with very small chemical potentials the QCD phase transition is a smooth transition \cite{Aoki:2006we} whereas the EW phase transition could be first-order in certain Standard Model extensions~\cite{Laine:1998qk}.

The conservation of magnetic helicity in the early Universe leads to an inverse cascade of energy and the slowing down of magnetic decay for fully helical fields. This means that, at the time when CMB spectral distortions can be generated, the magnetic field amplitude is relatively large compared to the non-helical case. This can lead to the generation of larger spectral distortions. If CMB spectral distortions are observed by some new PIXIE-like experiment, then it is likely that magnetic helicity plays an important role. However, there is a degeneracy in the parameter space, since different parameter sets can give the same spectral distortions signal. For example, fields generated at the QCD phase transition with smaller \m{\varepsilon\equiv\tilde{\rho}_{B,*}/\rho_{\gamma,0}} and/or helicity can produce the same $\mu$-type distortions as fields generated at the EW phase transition but with larger $\varepsilon$ and/or helicity. However, if a $\mu$-type distortion is detected by a PIXIE-like experiment, it would rule out non-helical magnetic fields produced at either the EW or QCD phase transition. A positive detection would give us a lower bound on the primordial magnetic helicity. The lower bound would be somewhere of the order \m{f_*\gtrsim(10^{-4}-10^{-3})}. This is much greater than the primordial magnetic helicity generated in the simplest models of EW baryogenesis \cite{Vachaspati:2001nb} where \m{f_*\sim10^{-24}} assuming \m{B_{\lambda,*}=B_{\lambda,*}^{\mathrm{max}}} and \m{\lambda_{B,*}=\lambda_{\mathrm{EW}}} \cite{Wagstaff:2014fla}. However, there are new mechanisms being proposed recently which can excite magnetic helicity in the early Universe due to the Chiral anomaly \cite{Boyarsky:2011uy,Boyarsky:2015faa}. It will be interesting to investigate such mechanisms in the future.

It is also interesting to mention the recent tentative observations of large scale helical magnetic fields from $\gamma$-ray observations \cite{Tashiro:2013ita}. Such studies have seen some evidence, albeit rather weak, of fully helical fields of strength \m{10^{-14}}~G on scales of \m{10}~Mpc. If such fields originated from a time before the $\mu$-era, then it is possible that such observations would be accompanied by a detectable signal for a PIXIE-like experiment. The combination of such two observations would be compelling evidence for large scale helical magnetic fields. 
We also note that the CMB distortions anisotropies (see e.g. Refs.~\cite{Miyamoto:2013oua,Ganc:2014wia}), albeit potentially very hard to detect, could give interesting signals due to the large helicity of the magnetic fields. Unique signatures in the spatial correlations are expected due to the helical nature of the magnetic fields. This will be investigated in future publications.

%%---------------------------------------------------------
%%---------------------------------------------------------

%\begin{acknowledgments}
\section*{Acknowledgments}
This work was supported by the Deutsche Forschungsgemeinschaft (DFG) through the collaborative research centre SFB 676 Particles, Strings, and the Early Universe, project C9.
%\end{acknowledgments}

%\end{widetext}

\bibliography{references_Mag}

%Merlin.mbs v4.21 2009-07-09.
\begin{thebibliography}{10}%
\makeatletter
\providecommand \@ifxundefined [1]{%
 \ifx #1\undefined \expandafter \@firstoftwo
 \else \expandafter \@secondoftwo
\fi
}%
\providecommand \@ifnum [1]{%
 \ifnum #1\expandafter \@firstoftwo
 \else \expandafter \@secondoftwo
\fi
}%
\providecommand \enquote [1]{``#1''}%
\providecommand \bibnamefont  [1]{#1}%
\providecommand \bibfnamefont [1]{#1}%
\providecommand \citenamefont [1]{#1}%
\providecommand\href[0]{\@sanitize\@href}%
\providecommand\@href[1]{\endgroup\@@startlink{#1}\endgroup\@@href}%
\providecommand\@@href[1]{#1\@@endlink}%
\providecommand \@sanitize [0]{\begingroup\catcode`\&12\catcode`\#12\relax}%
\@ifxundefined \pdfoutput {\@firstoftwo}{%
 \@ifnum{\z@=\pdfoutput}{\@firstoftwo}{\@secondoftwo}%
}{%
 \providecommand\@@startlink[1]{\leavevmode\special{html:<a href="#1">}}%
 \providecommand\@@endlink[0]{\special{html:</a>}}%
}{%
 \providecommand\@@startlink[1]{%
  \leavevmode
  \pdfstartlink
   attr{/Border[0 0 1 ]/H/I/C[0 1 1]}%
   user{/Subtype/Link/A<</Type/Action/S/URI/URI(#1)>>}%
  \relax
 }%
 \providecommand\@@endlink[0]{\pdfendlink}%
}%
\providecommand \url  [0]{\begingroup\@sanitize \@url }%
\providecommand \@url [1]{\endgroup\@href {#1}{\urlprefix}}%
\providecommand \urlprefix [0]{URL }%
\providecommand \Eprint[0]{\href }%
\@ifxundefined \urlstyle {%
  \providecommand \doi [1]{doi:\discretionary{}{}{}#1}%
}{%
  \providecommand \doi [0]{doi:\discretionary{}{}{}\begingroup
  \urlstyle{rm}\Url }%
}%
\providecommand \doibase [0]{http://dx.doi.org/}%
\providecommand \Doi[1]{\href{\doibase#1}}%
\providecommand \bibAnnote [3]{%
  \BibitemShut{#1}%
  \begin{quotation}\noindent
    \textsc{Key:}\ #2\\\textsc{Annotation:}\ #3%
  \end{quotation}%
}%
\providecommand \bibAnnoteFile [2]{%
  \IfFileExists{#2}{\bibAnnote {#1} {#2} {\input{#2}}}{}%
}%
\providecommand \typeout [0]{\immediate \write \m@ne }%
\providecommand \selectlanguage [0]{\@gobble}%
\providecommand \bibinfo [0]{\@secondoftwo}%
\providecommand \bibfield [0]{\@secondoftwo}%
\providecommand \translation [1]{[#1]}%
\providecommand \BibitemOpen[0]{}%
\providecommand \bibitemStop [0]{}%
\providecommand \bibitemNoStop [0]{.\EOS\space}%
\providecommand \EOS [0]{\spacefactor3000\relax}%
\providecommand \BibitemShut [1]{\csname bibitem#1\endcsname}%
%</preamble>
\bibitem{BFields_galax}%
  \BibitemOpen
  \bibfield{author}{%
  \bibinfo {author} {\bibfnamefont{R.}~\bibnamefont{Beck}},\ }%
  \bibfield{journal}{%
  \Doi{10.1007/s11214-011-9782-z}{\bibinfo {journal} {Space Science Reviews}}\
  }%
  \textbf{\bibinfo {volume} {166}},\ \bibinfo {pages} {215} (\bibinfo {year}
  {2012})%
  \bibAnnoteFile{NoStop}{BFields_galax}%
\bibitem{BFields_Bernet}%
  \BibitemOpen
  \bibfield{author}{%
  \bibinfo {author} {\bibfnamefont{M.~L.}\ \bibnamefont{Bernet}}, \bibinfo
  {author} {\bibfnamefont{F.}~\bibnamefont{Miniati}}, \bibinfo {author}
  {\bibfnamefont{S.~J.}\ \bibnamefont{Lilly}}, \bibinfo {author}
  {\bibfnamefont{P.~P.}\ \bibnamefont{Kronberg}},\ and\ \bibinfo {author}
  {\bibfnamefont{M.}~\bibnamefont{Dessauges-Zavadsky}},\ }%
  \bibfield{journal}{%
  \Doi{10.1038/nature07105}{\bibinfo {journal} {Nature}}\ }%
  \textbf{\bibinfo {volume} {454}},\ \bibinfo {pages} {302} (\bibinfo {year}
  {2008}),\ \Eprint{http://arxiv.org/abs/0807.3347}{arXiv:0807.3347}%
  \bibAnnoteFile{NoStop}{BFields_Bernet}%
\bibitem{greenpeas}%
  \BibitemOpen
  \bibfield{author}{%
  \bibinfo {author} {\bibfnamefont{S.}~\bibnamefont{{Chakraborti}}}, \bibinfo
  {author} {\bibfnamefont{N.}~\bibnamefont{{Yadav}}}, \bibinfo {author}
  {\bibfnamefont{C.}~\bibnamefont{{Cardamone}}},\ and\ \bibinfo {author}
  {\bibfnamefont{A.}~\bibnamefont{{Ray}}},\ }%
  \bibfield{journal}{%
  \Doi{10.1088/2041-8205/746/1/L6}{\bibinfo {journal} {Astrophys.\, J.\,}}\ }%
  \textbf{\bibinfo {volume} {746}},\ \bibinfo {eid} {L6} (\bibinfo {month}
  {Feb.}\ \bibinfo {year} {2012}),\
  \Eprint{http://arxiv.org/abs/1110.3312}{arXiv:1110.3312 [astro-ph.CO]}%
  \bibAnnoteFile{NoStop}{greenpeas}%
\bibitem{BFields-Clusters}%
  \BibitemOpen
  \bibfield{author}{%
  \bibinfo {author} {\bibfnamefont{L.}~\bibnamefont{Feretti}}, \bibinfo
  {author} {\bibfnamefont{G.}~\bibnamefont{Giovannini}}, \bibinfo {author}
  {\bibfnamefont{F.}~\bibnamefont{Govoni}},\ and\ \bibinfo {author}
  {\bibfnamefont{M.}~\bibnamefont{Murgia}},\ }%
  \bibfield{journal}{%
  \Doi{10.1007/s00159-012-0054-z}{\bibinfo {journal} {A\&AR}}\ }%
  \textbf{\bibinfo {volume} {20}},\ \bibinfo {pages} {54} (\bibinfo {year}
  {2012}),\ \Eprint{http://arxiv.org/abs/1205.1919}{arXiv:1205.1919
  [astro-ph.CO]}%
  \bibAnnoteFile{NoStop}{BFields-Clusters}%
\bibitem{BFields_SClusters}%
  \BibitemOpen
  \bibfield{author}{%
  \bibinfo {author} {\bibfnamefont{Y.}~\bibnamefont{Xu}}, \bibinfo {author}
  {\bibfnamefont{P.~P.}\ \bibnamefont{Kronberg}}, \bibinfo {author}
  {\bibfnamefont{S.}~\bibnamefont{Habib}},\ and\ \bibinfo {author}
  {\bibfnamefont{Q.~W.}\ \bibnamefont{Dufton}},\ }%
  \bibfield{journal}{%
  \Doi{10.1086/498336}{\bibinfo {journal} {AstrophysJ}}\ }%
  \textbf{\bibinfo {volume} {637}},\ \bibinfo {pages} {19} (\bibinfo {year}
  {2006}),\
  \Eprint{http://arxiv.org/abs/arXiv:astro-ph/0509826}{arXiv:astro-ph/0509826}%
  \bibAnnoteFile{NoStop}{BFields_SClusters}%
\bibitem{Neronov:1900zz}%
  \BibitemOpen
  \bibfield{author}{%
  \bibinfo {author} {\bibfnamefont{A.}~\bibnamefont{Neronov}}\ and\ \bibinfo
  {author} {\bibfnamefont{I.}~\bibnamefont{Vovk}},\ }%
  \bibfield{journal}{%
  \Doi{10.1126/science.1184192}{\bibinfo {journal} {Science}}\ }%
  \textbf{\bibinfo {volume} {328}},\ \bibinfo {pages} {73} (\bibinfo {year}
  {2010}),\ \Eprint{http://arxiv.org/abs/1006.3504}{arXiv:1006.3504
  [astro-ph.HE]}%
  \bibAnnoteFile{NoStop}{Neronov:1900zz}%
%%CITATION = ARXIV:1006.3504;%%
\bibitem{Turner:PMF}%
  \BibitemOpen
  \bibfield{author}{%
  \bibinfo {author} {\bibfnamefont{M.~S.}\ \bibnamefont{Turner}}\ and\ \bibinfo
  {author} {\bibfnamefont{L.~M.}\ \bibnamefont{Widrow}},\ }%
  \bibfield{journal}{%
  \Doi{10.1103/PhysRevD.37.2743}{\bibinfo {journal} {Phys.Rev.}}\ }%
  \textbf{\bibinfo {volume} {D37}},\ \bibinfo {pages} {2743} (\bibinfo {year}
  {1988})%
  \bibAnnoteFile{NoStop}{Turner:PMF}%
%%CITATION = PHRVA,D37,2743;%%
\bibitem{B-phaseT}%
  \BibitemOpen
  \bibfield{author}{%
  \bibinfo {author} {\bibfnamefont{G.}~\bibnamefont{Sigl}}, \bibinfo {author}
  {\bibfnamefont{A.~V.}\ \bibnamefont{Olinto}},\ and\ \bibinfo {author}
  {\bibfnamefont{K.}~\bibnamefont{Jedamzik}},\ }%
  \bibfield{journal}{%
  \Doi{10.1103/PhysRevD.55.4582}{\bibinfo {journal} {Phys.Rev.}}\ }%
  \textbf{\bibinfo {volume} {D55}},\ \bibinfo {pages} {4582} (\bibinfo {year}
  {1997}),\
  \Eprint{http://arxiv.org/abs/astro-ph/9610201}{arXiv:astro-ph/9610201
  [astro-ph]}%
  \bibAnnoteFile{NoStop}{B-phaseT}%
%%CITATION = ASTRO-PH/9610201;%%
\bibitem{Harrison}%
  \BibitemOpen
  \bibfield{author}{%
  \bibinfo {author} {\bibfnamefont{E.}~\bibnamefont{Harrison}},\ }%
  \bibfield{journal}{%
  \bibinfo {journal} {Mon.Not.Roy.Astron.Soc.}\ }%
  \textbf{\bibinfo {volume} {147}},\ \bibinfo {pages} {279} (\bibinfo {year}
  {1970})%
  \bibAnnoteFile{NoStop}{Harrison}%
\bibitem{Dolgov:2001nv}%
  \BibitemOpen
  \bibfield{author}{%
  \bibinfo {author} {\bibfnamefont{A.~D.}\ \bibnamefont{Dolgov}}\ and\ \bibinfo
  {author} {\bibfnamefont{D.}~\bibnamefont{Grasso}},\ }%
  \bibfield{journal}{%
  \Doi{10.1103/PhysRevLett.88.011301}{\bibinfo {journal} {Phys.Rev.Lett.}}\ }%
  \textbf{\bibinfo {volume} {88}},\ \bibinfo {pages} {011301} (\bibinfo {year}
  {2001}),\
  \Eprint{http://arxiv.org/abs/astro-ph/0106154}{arXiv:astro-ph/0106154
  [astro-ph]}%
  \bibAnnoteFile{NoStop}{Dolgov:2001nv}%
%%CITATION = ASTRO-PH/0106154;%%
\bibitem{Christensson:2000sp}%
  \BibitemOpen
  \bibfield{author}{%
  \bibinfo {author} {\bibfnamefont{M.}~\bibnamefont{Christensson}}, \bibinfo
  {author} {\bibfnamefont{M.}~\bibnamefont{Hindmarsh}},\ and\ \bibinfo {author}
  {\bibfnamefont{A.}~\bibnamefont{Brandenburg}},\ }%
  \bibfield{journal}{%
  \Doi{10.1103/PhysRevE.64.056405}{\bibinfo {journal} {Phys.Rev.}}\ }%
  \textbf{\bibinfo {volume} {E64}},\ \bibinfo {pages} {056405} (\bibinfo {year}
  {2001}),\
  \Eprint{http://arxiv.org/abs/astro-ph/0011321}{arXiv:astro-ph/0011321
  [astro-ph]}%
  \bibAnnoteFile{NoStop}{Christensson:2000sp}%
%%CITATION = ASTRO-PH/0011321;%%
\bibitem{Banerjee:2004df}%
  \BibitemOpen
  \bibfield{author}{%
  \bibinfo {author} {\bibfnamefont{R.}~\bibnamefont{Banerjee}}\ and\ \bibinfo
  {author} {\bibfnamefont{K.}~\bibnamefont{Jedamzik}},\ }%
  \bibfield{journal}{%
  \Doi{10.1103/PhysRevD.70.123003}{\bibinfo {journal} {Phys.Rev.}}\ }%
  \textbf{\bibinfo {volume} {D70}},\ \bibinfo {pages} {123003} (\bibinfo {year}
  {2004}),\
  \Eprint{http://arxiv.org/abs/astro-ph/0410032}{arXiv:astro-ph/0410032
  [astro-ph]}%
  \bibAnnoteFile{NoStop}{Banerjee:2004df}%
%%CITATION = ASTRO-PH/0410032;%%
\bibitem{Campanelli:2013iaa}%
  \BibitemOpen
  \bibfield{author}{%
  \bibinfo {author} {\bibfnamefont{L.}~\bibnamefont{Campanelli}},\ }%
  \bibfield{journal}{%
  \Doi{10.1140/epjc/s10052-013-2690-5}{\bibinfo {journal} {Eur.Phys.J.}}\ }%
  \textbf{\bibinfo {volume} {C74}},\ \bibinfo {pages} {2690} (\bibinfo {year}
  {2014}),\ \Eprint{http://arxiv.org/abs/1304.4044}{arXiv:1304.4044
  [astro-ph.CO]}%
  \bibAnnoteFile{NoStop}{Campanelli:2013iaa}%
%%CITATION = ARXIV:1304.4044;%%
\bibitem{CMB_BB}%
  \BibitemOpen
  \bibfield{author}{%
  \bibinfo {author} {\bibfnamefont{D.}~\bibnamefont{Fixsen}}, \bibinfo {author}
  {\bibfnamefont{E.}~\bibnamefont{Cheng}}, \bibinfo {author}
  {\bibfnamefont{J.}~\bibnamefont{Gales}}, \bibinfo {author}
  {\bibfnamefont{J.~C.}\ \bibnamefont{Mather}}, \bibinfo {author}
  {\bibfnamefont{R.}~\bibnamefont{Shafer}}, \emph{et~al.},\ }%
  \bibfield{journal}{%
  \Doi{10.1086/178173}{\bibinfo {journal} {Astrophys.J.}}\ }%
  \textbf{\bibinfo {volume} {473}},\ \bibinfo {pages} {576} (\bibinfo {year}
  {1996}),\
  \Eprint{http://arxiv.org/abs/astro-ph/9605054}{arXiv:astro-ph/9605054
  [astro-ph]}%
  \bibAnnoteFile{NoStop}{CMB_BB}%
%%CITATION = ASTRO-PH/9605054;%%
\bibitem{Sunyaev1969}%
  \BibitemOpen
  \bibfield{author}{%
  \bibinfo {author} {\bibfnamefont{R.~A.}\ \bibnamefont{{Sunyaev}}}\ and\
  \bibinfo {author} {\bibfnamefont{Y.~B.}\ \bibnamefont{{Zeldovich}}},\ }%
  \bibfield{journal}{%
  \Doi{10.1038/223721a0}{\bibinfo {journal} {\nat}}\ }%
  \textbf{\bibinfo {volume} {223}},\ \bibinfo {pages} {721} (\bibinfo {month}
  {Aug.}\ \bibinfo {year} {1969})%
  \bibAnnoteFile{NoStop}{Sunyaev1969}%
\bibitem{HuSilk:1993}%
  \BibitemOpen
  \bibfield{author}{%
  \bibinfo {author} {\bibfnamefont{W.}~\bibnamefont{Hu}}\ and\ \bibinfo
  {author} {\bibfnamefont{J.}~\bibnamefont{Silk}},\ }%
  \bibfield{journal}{%
  \Doi{10.1103/PhysRevD.48.485}{\bibinfo {journal} {Phys.Rev.}}\ }%
  \textbf{\bibinfo {volume} {D48}},\ \bibinfo {pages} {485} (\bibinfo {year}
  {1993})%
  \bibAnnoteFile{NoStop}{HuSilk:1993}%
%%CITATION = PHRVA,D48,485;%%
\bibitem{Khatri:2012rt}%
  \BibitemOpen
  \bibfield{author}{%
  \bibinfo {author} {\bibfnamefont{R.}~\bibnamefont{Khatri}}, \bibinfo {author}
  {\bibfnamefont{R.~A.}\ \bibnamefont{Sunyaev}},\ and\ \bibinfo {author}
  {\bibfnamefont{J.}~\bibnamefont{Chluba}},\ }%
  \bibfield{journal}{%
  \Doi{10.1051/0004-6361/201219590}{\bibinfo {journal} {Astron.Astrophys.}}\ }%
  \textbf{\bibinfo {volume} {543}},\ \bibinfo {pages} {A136} (\bibinfo {year}
  {2012}),\ \Eprint{http://arxiv.org/abs/1205.2871}{arXiv:1205.2871
  [astro-ph.CO]}%
  \bibAnnoteFile{NoStop}{Khatri:2012rt}%
%%CITATION = ARXIV:1205.2871;%%
\bibitem{Chluba:2012gq}%
  \BibitemOpen
  \bibfield{author}{%
  \bibinfo {author} {\bibfnamefont{J.}~\bibnamefont{Chluba}}, \bibinfo {author}
  {\bibfnamefont{R.}~\bibnamefont{Khatri}},\ and\ \bibinfo {author}
  {\bibfnamefont{R.~A.}\ \bibnamefont{Sunyaev}},\ }%
  \bibfield{journal}{%
  \Doi{10.1111/j.1365-2966.2012.21474.x}{\bibinfo {journal}
  {Mon.Not.Roy.Astron.Soc.}}\ }%
  \textbf{\bibinfo {volume} {425}},\ \bibinfo {pages} {1129} (\bibinfo {year}
  {2012}),\ \Eprint{http://arxiv.org/abs/1202.0057}{arXiv:1202.0057
  [astro-ph.CO]}%
  \bibAnnoteFile{NoStop}{Chluba:2012gq}%
%%CITATION = ARXIV:1202.0057;%%
\bibitem{Pajer:2013oca}%
  \BibitemOpen
  \bibfield{author}{%
  \bibinfo {author} {\bibfnamefont{E.}~\bibnamefont{Pajer}}\ and\ \bibinfo
  {author} {\bibfnamefont{M.}~\bibnamefont{Zaldarriaga}},\ }%
  \bibfield{journal}{%
  \Doi{10.1088/1475-7516/2013/02/036}{\bibinfo {journal} {JCAP}}\ }%
  \textbf{\bibinfo {volume} {1302}},\ \bibinfo {pages} {036} (\bibinfo {year}
  {2013}),\ \Eprint{http://arxiv.org/abs/1206.4479}{arXiv:1206.4479
  [astro-ph.CO]}%
  \bibAnnoteFile{NoStop}{Pajer:2013oca}%
%%CITATION = ARXIV:1206.4479;%%
\bibitem{KK14}%
  \BibitemOpen
  \bibfield{author}{%
  \bibinfo {author} {\bibfnamefont{K.~E.}\ \bibnamefont{Kunze}}\ and\ \bibinfo
  {author} {\bibfnamefont{E.}~\bibnamefont{Komatsu}},\ }%
  \bibfield{journal}{%
  \Doi{10.1088/1475-7516/2014/01/009}{\bibinfo {journal} {JCAP}}\ }%
  \textbf{\bibinfo {volume} {1401}},\ \bibinfo {pages} {009} (\bibinfo {year}
  {2014}),\ \Eprint{http://arxiv.org/abs/1309.7994}{arXiv:1309.7994
  [astro-ph.CO]}%
  \bibAnnoteFile{NoStop}{KK14}%
%%CITATION = ARXIV:1309.7994;%%
\bibitem{Jedamzik_CMBdist}%
  \BibitemOpen
  \bibfield{author}{%
  \bibinfo {author} {\bibfnamefont{K.}~\bibnamefont{Jedamzik}}, \bibinfo
  {author} {\bibfnamefont{V.}~\bibnamefont{Katalinic}},\ and\ \bibinfo {author}
  {\bibfnamefont{A.~V.}\ \bibnamefont{Olinto}},\ }%
  \bibfield{journal}{%
  \Doi{10.1103/PhysRevLett.85.700}{\bibinfo {journal} {Phys.Rev.Lett.}}\ }%
  \textbf{\bibinfo {volume} {85}},\ \bibinfo {pages} {700} (\bibinfo {year}
  {2000}),\
  \Eprint{http://arxiv.org/abs/astro-ph/9911100}{arXiv:astro-ph/9911100
  [astro-ph]}%
  \bibAnnoteFile{NoStop}{Jedamzik_CMBdist}%
%%CITATION = ASTRO-PH/9911100;%%
\bibitem{Miyamoto:2013oua}%
  \BibitemOpen
  \bibfield{author}{%
  \bibinfo {author} {\bibfnamefont{K.}~\bibnamefont{Miyamoto}}, \bibinfo
  {author} {\bibfnamefont{T.}~\bibnamefont{Sekiguchi}}, \bibinfo {author}
  {\bibfnamefont{H.}~\bibnamefont{Tashiro}},\ and\ \bibinfo {author}
  {\bibfnamefont{S.}~\bibnamefont{Yokoyama}},\ }%
  \bibfield{journal}{%
  \Doi{10.1103/PhysRevD.89.063508}{\bibinfo {journal} {Phys. Rev.}}\ }%
  \textbf{\bibinfo {volume} {D89}},\ \bibinfo {pages} {063508} (\bibinfo {year}
  {2014}),\ \Eprint{http://arxiv.org/abs/1310.3886}{arXiv:1310.3886
  [astro-ph.CO]}%
  \bibAnnoteFile{NoStop}{Miyamoto:2013oua}%
%%CITATION = ARXIV:1310.3886;%%
\bibitem{Amin:2014ada}%
  \BibitemOpen
  \bibfield{author}{%
  \bibinfo {author} {\bibfnamefont{M.~A.}\ \bibnamefont{Amin}}\ and\ \bibinfo
  {author} {\bibfnamefont{D.}~\bibnamefont{Grin}},\ }%
  \bibfield{journal}{%
  \Doi{10.1103/PhysRevD.90.083529}{\bibinfo {journal} {Phys. Rev.}}\ }%
  \textbf{\bibinfo {volume} {D90}},\ \bibinfo {pages} {083529} (\bibinfo {year}
  {2014}),\ \Eprint{http://arxiv.org/abs/1405.1039}{arXiv:1405.1039
  [astro-ph.CO]}%
  \bibAnnoteFile{NoStop}{Amin:2014ada}%
%%CITATION = ARXIV:1405.1039;%%
\bibitem{Tashiro:2014pga}%
  \BibitemOpen
  \bibfield{author}{%
  \bibinfo {author} {\bibfnamefont{H.}~\bibnamefont{Tashiro}},\ }%
  \bibfield{journal}{%
  \Doi{10.1093/ptep/ptu066}{\bibinfo {journal} {PTEP}}\ }%
  \textbf{\bibinfo {volume} {2014}},\ \bibinfo {pages} {06B107} (\bibinfo
  {year} {2014})%
  \bibAnnoteFile{NoStop}{Tashiro:2014pga}%
%%CITATION = INSPIRE-1301471;%%
\bibitem{Khatri:2012tw}%
  \BibitemOpen
  \bibfield{author}{%
  \bibinfo {author} {\bibfnamefont{R.}~\bibnamefont{Khatri}}\ and\ \bibinfo
  {author} {\bibfnamefont{R.~A.}\ \bibnamefont{Sunyaev}},\ }%
  \bibfield{journal}{%
  \Doi{10.1088/1475-7516/2012/09/016}{\bibinfo {journal} {JCAP}}\ }%
  \textbf{\bibinfo {volume} {1209}},\ \bibinfo {pages} {016} (\bibinfo {year}
  {2012}),\ \Eprint{http://arxiv.org/abs/1207.6654}{arXiv:1207.6654
  [astro-ph.CO]}%
  \bibAnnoteFile{NoStop}{Khatri:2012tw}%
%%CITATION = ARXIV:1207.6654;%%
\bibitem{Hu:1994bz}%
  \BibitemOpen
  \bibfield{author}{%
  \bibinfo {author} {\bibfnamefont{W.}~\bibnamefont{Hu}}, \bibinfo {author}
  {\bibfnamefont{D.}~\bibnamefont{Scott}},\ and\ \bibinfo {author}
  {\bibfnamefont{J.}~\bibnamefont{Silk}},\ }%
  \bibfield{journal}{%
  \Doi{10.1086/187424}{\bibinfo {journal} {Astrophys. J.}}\ }%
  \textbf{\bibinfo {volume} {430}},\ \bibinfo {pages} {L5} (\bibinfo {year}
  {1994}),\
  \Eprint{http://arxiv.org/abs/astro-ph/9402045}{arXiv:astro-ph/9402045
  [astro-ph]}%
  \bibAnnoteFile{NoStop}{Hu:1994bz}%
%%CITATION = ASTRO-PH/9402045;%%
\bibitem{McDonald:2000bk}%
  \BibitemOpen
  \bibfield{author}{%
  \bibinfo {author} {\bibfnamefont{P.}~\bibnamefont{McDonald}}, \bibinfo
  {author} {\bibfnamefont{R.~J.}\ \bibnamefont{Scherrer}},\ and\ \bibinfo
  {author} {\bibfnamefont{T.~P.}\ \bibnamefont{Walker}},\ }%
  \bibfield{journal}{%
  \Doi{10.1103/PhysRevD.63.023001}{\bibinfo {journal} {Phys. Rev.}}\ }%
  \textbf{\bibinfo {volume} {D63}},\ \bibinfo {pages} {023001} (\bibinfo {year}
  {2001}),\
  \Eprint{http://arxiv.org/abs/astro-ph/0008134}{arXiv:astro-ph/0008134
  [astro-ph]}%
  \bibAnnoteFile{NoStop}{McDonald:2000bk}%
%%CITATION = ASTRO-PH/0008134;%%
\bibitem{Chluba:2011hw}%
  \BibitemOpen
  \bibfield{author}{%
  \bibinfo {author} {\bibfnamefont{J.}~\bibnamefont{Chluba}}\ and\ \bibinfo
  {author} {\bibfnamefont{R.~A.}\ \bibnamefont{Sunyaev}},\ }%
  \bibfield{journal}{%
  \Doi{10.1111/j.1365-2966.2011.19786.x}{\bibinfo {journal} {Mon. Not. Roy.
  Astron. Soc.}}\ }%
  \textbf{\bibinfo {volume} {419}},\ \bibinfo {pages} {1294} (\bibinfo {year}
  {2012}),\ \Eprint{http://arxiv.org/abs/1109.6552}{arXiv:1109.6552
  [astro-ph.CO]}%
  \bibAnnoteFile{NoStop}{Chluba:2011hw}%
%%CITATION = ARXIV:1109.6552;%%
\bibitem{Chluba:2013wsa}%
  \BibitemOpen
  \bibfield{author}{%
  \bibinfo {author} {\bibfnamefont{J.}~\bibnamefont{Chluba}},\ }%
  \bibfield{journal}{%
  \Doi{10.1093/mnras/stt1733}{\bibinfo {journal} {Mon.Not.Roy.Astron.Soc.}}\ }%
  \textbf{\bibinfo {volume} {436}},\ \bibinfo {pages} {2232} (\bibinfo {year}
  {2013}),\ \Eprint{http://arxiv.org/abs/1304.6121}{arXiv:1304.6121
  [astro-ph.CO]}%
  \bibAnnoteFile{NoStop}{Chluba:2013wsa}%
%%CITATION = ARXIV:1304.6121;%%
\bibitem{Carr:2009jm}%
  \BibitemOpen
  \bibfield{author}{%
  \bibinfo {author} {\bibfnamefont{B.~J.}\ \bibnamefont{Carr}}, \bibinfo
  {author} {\bibfnamefont{K.}~\bibnamefont{Kohri}}, \bibinfo {author}
  {\bibfnamefont{Y.}~\bibnamefont{Sendouda}},\ and\ \bibinfo {author}
  {\bibfnamefont{J.}~\bibnamefont{Yokoyama}},\ }%
  \bibfield{journal}{%
  \Doi{10.1103/PhysRevD.81.104019}{\bibinfo {journal} {Phys. Rev.}}\ }%
  \textbf{\bibinfo {volume} {D81}},\ \bibinfo {pages} {104019} (\bibinfo {year}
  {2010}),\ \Eprint{http://arxiv.org/abs/0912.5297}{arXiv:0912.5297
  [astro-ph.CO]}%
  \bibAnnoteFile{NoStop}{Carr:2009jm}%
%%CITATION = ARXIV:0912.5297;%%
\bibitem{Tashiro:2012pp}%
  \BibitemOpen
  \bibfield{author}{%
  \bibinfo {author} {\bibfnamefont{H.}~\bibnamefont{Tashiro}}, \bibinfo
  {author} {\bibfnamefont{E.}~\bibnamefont{Sabancilar}},\ and\ \bibinfo
  {author} {\bibfnamefont{T.}~\bibnamefont{Vachaspati}},\ }%
  \bibfield{journal}{%
  \Doi{10.1088/1475-7516/2013/08/035}{\bibinfo {journal} {JCAP}}\ }%
  \textbf{\bibinfo {volume} {1308}},\ \bibinfo {pages} {035} (\bibinfo {year}
  {2013}),\ \Eprint{http://arxiv.org/abs/1212.3283}{arXiv:1212.3283
  [astro-ph.CO]}%
  \bibAnnoteFile{NoStop}{Tashiro:2012pp}%
%%CITATION = ARXIV:1212.3283;%%
\bibitem{CausalBfield}%
  \BibitemOpen
  \bibfield{author}{%
  \bibinfo {author} {\bibfnamefont{R.}~\bibnamefont{Durrer}}\ and\ \bibinfo
  {author} {\bibfnamefont{C.}~\bibnamefont{Caprini}},\ }%
  \bibfield{journal}{%
  \Doi{10.1088/1475-7516/2003/11/010}{\bibinfo {journal} {JCAP}}\ }%
  \textbf{\bibinfo {volume} {0311}},\ \bibinfo {pages} {010} (\bibinfo {year}
  {2003}),\
  \Eprint{http://arxiv.org/abs/astro-ph/0305059}{arXiv:astro-ph/0305059
  [astro-ph]}%
  \bibAnnoteFile{NoStop}{CausalBfield}%
%%CITATION = ASTRO-PH/0305059;%%
\bibitem{Campanelli:2007tc}%
  \BibitemOpen
  \bibfield{author}{%
  \bibinfo {author} {\bibfnamefont{L.}~\bibnamefont{Campanelli}},\ }%
  \bibfield{journal}{%
  \Doi{10.1103/PhysRevLett.98.251302}{\bibinfo {journal} {Phys.Rev.Lett.}}\ }%
  \textbf{\bibinfo {volume} {98}},\ \bibinfo {pages} {251302} (\bibinfo {year}
  {2007}),\ \Eprint{http://arxiv.org/abs/0705.2308}{arXiv:0705.2308
  [astro-ph]}%
  \bibAnnoteFile{NoStop}{Campanelli:2007tc}%
%%CITATION = ARXIV:0705.2308;%%
\bibitem{Saveliev:2012ea}%
  \BibitemOpen
  \bibfield{author}{%
  \bibinfo {author} {\bibfnamefont{A.}~\bibnamefont{Saveliev}}, \bibinfo
  {author} {\bibfnamefont{K.}~\bibnamefont{Jedamzik}},\ and\ \bibinfo {author}
  {\bibfnamefont{G.}~\bibnamefont{Sigl}},\ }%
  \bibfield{journal}{%
  \Doi{10.1103/PhysRevD.86.103010}{\bibinfo {journal} {Phys.Rev.}}\ }%
  \textbf{\bibinfo {volume} {D86}},\ \bibinfo {pages} {103010} (\bibinfo {year}
  {2012}),\ \Eprint{http://arxiv.org/abs/1208.0444}{arXiv:1208.0444
  [astro-ph.CO]}%
  \bibAnnoteFile{NoStop}{Saveliev:2012ea}%
%%CITATION = ARXIV:1208.0444;%%
\bibitem{Saveliev:2013uva}%
  \BibitemOpen
  \bibfield{author}{%
  \bibinfo {author} {\bibfnamefont{A.}~\bibnamefont{Saveliev}}, \bibinfo
  {author} {\bibfnamefont{K.}~\bibnamefont{Jedamzik}},\ and\ \bibinfo {author}
  {\bibfnamefont{G.}~\bibnamefont{Sigl}},\ }%
  \bibfield{journal}{%
  \Doi{10.1103/PhysRevD.87.123001}{\bibinfo {journal} {Phys.Rev.}}\ }%
  \textbf{\bibinfo {volume} {D87}},\ \bibinfo {pages} {123001} (\bibinfo {year}
  {2013}),\ \Eprint{http://arxiv.org/abs/1304.3621}{arXiv:1304.3621
  [astro-ph.CO]}%
  \bibAnnoteFile{NoStop}{Saveliev:2013uva}%
%%CITATION = ARXIV:1304.3621;%%
\bibitem{Brandenburg:2014mwa}%
  \BibitemOpen
  \bibfield{author}{%
  \bibinfo {author} {\bibfnamefont{A.}~\bibnamefont{Brandenburg}}, \bibinfo
  {author} {\bibfnamefont{T.}~\bibnamefont{Kahniashvili}},\ and\ \bibinfo
  {author} {\bibfnamefont{A.~G.}\ \bibnamefont{Tevzadze}},\ }%
  \bibfield{journal}{%
  \Doi{10.1103/PhysRevLett.114.075001}{\bibinfo {journal} {Phys.Rev.Lett.}}\ }%
  \textbf{\bibinfo {volume} {114}},\ \bibinfo {pages} {075001} (\bibinfo {year}
  {2015}),\ \Eprint{http://arxiv.org/abs/1404.2238}{arXiv:1404.2238
  [astro-ph.CO]}%
  \bibAnnoteFile{NoStop}{Brandenburg:2014mwa}%
%%CITATION = ARXIV:1404.2238;%%
\bibitem{Durrer:2013pga}%
  \BibitemOpen
  \bibfield{author}{%
  \bibinfo {author} {\bibfnamefont{R.}~\bibnamefont{Durrer}}\ and\ \bibinfo
  {author} {\bibfnamefont{A.}~\bibnamefont{Neronov}},\ }%
  \bibfield{journal}{%
  \Doi{10.1007/s00159-013-0062-7}{\bibinfo {journal} {Astron.Astrophys.Rev.}}\
  }%
  \textbf{\bibinfo {volume} {21}},\ \bibinfo {pages} {62} (\bibinfo {year}
  {2013}),\ \Eprint{http://arxiv.org/abs/1303.7121}{arXiv:1303.7121
  [astro-ph.CO]}%
  \bibAnnoteFile{NoStop}{Durrer:2013pga}%
%%CITATION = ARXIV:1303.7121;%%
\bibitem{Campanelli_free-turb}%
  \BibitemOpen
  \bibfield{author}{%
  \bibinfo {author} {\bibfnamefont{L.}~\bibnamefont{Campanelli}},\ }%
  \bibfield{journal}{%
  \Doi{10.1103/PhysRevLett.98.251302}{\bibinfo {journal} {Phys.Rev.Lett.}}\ }%
  \textbf{\bibinfo {volume} {98}},\ \bibinfo {pages} {251302} (\bibinfo {year}
  {2007}),\ \Eprint{http://arxiv.org/abs/0705.2308}{arXiv:0705.2308
  [astro-ph]}%
  \bibAnnoteFile{NoStop}{Campanelli_free-turb}%
%%CITATION = ARXIV:0705.2308;%%
\bibitem{Wagstaff:2013yna}%
  \BibitemOpen
  \bibfield{author}{%
  \bibinfo {author} {\bibfnamefont{J.~M.}\ \bibnamefont{Wagstaff}}, \bibinfo
  {author} {\bibfnamefont{R.}~\bibnamefont{Banerjee}}, \bibinfo {author}
  {\bibfnamefont{D.}~\bibnamefont{Schleicher}},\ and\ \bibinfo {author}
  {\bibfnamefont{G.}~\bibnamefont{Sigl}},\ }%
  \bibfield{journal}{%
  \Doi{10.1103/PhysRevD.89.103001}{\bibinfo {journal} {Phys.Rev.}}\ }%
  \textbf{\bibinfo {volume} {D89}},\ \bibinfo {pages} {103001} (\bibinfo {year}
  {2014}),\ \Eprint{http://arxiv.org/abs/1304.4723}{arXiv:1304.4723
  [astro-ph.CO]}%
  \bibAnnoteFile{NoStop}{Wagstaff:2013yna}%
%%CITATION = ARXIV:1304.4723;%%
\bibitem{Biskamp1993}%
  \BibitemOpen
  \bibfield{author}{%
  \bibinfo {author} {\bibfnamefont{D.}~\bibnamefont{{Biskamp}}},\ }%
  \emph{\bibinfo {title} {Cambridge Monographs on Plasma Physics, Cambridge
  [England]; New York, NY: Cambridge University Press, |c1993}}\ (\bibinfo
  {year} {1993})%
  \bibAnnoteFile{NoStop}{Biskamp1993}%
\bibitem{Kahniashvili:2012uj}%
  \BibitemOpen
  \bibfield{author}{%
  \bibinfo {author} {\bibfnamefont{T.}~\bibnamefont{Kahniashvili}}, \bibinfo
  {author} {\bibfnamefont{A.~G.}\ \bibnamefont{Tevzadze}}, \bibinfo {author}
  {\bibfnamefont{A.}~\bibnamefont{Brandenburg}},\ and\ \bibinfo {author}
  {\bibfnamefont{A.}~\bibnamefont{Neronov}},\ }%
  \bibfield{journal}{%
  \Doi{10.1103/PhysRevD.87.083007}{\bibinfo {journal} {Phys.Rev.}}\ }%
  \textbf{\bibinfo {volume} {D87}},\ \bibinfo {pages} {083007} (\bibinfo {year}
  {2013}),\ \Eprint{http://arxiv.org/abs/1212.0596}{arXiv:1212.0596
  [astro-ph.CO]}%
  \bibAnnoteFile{NoStop}{Kahniashvili:2012uj}%
%%CITATION = ARXIV:1212.0596;%%
\bibitem{Campanelli:2004wm}%
  \BibitemOpen
  \bibfield{author}{%
  \bibinfo {author} {\bibfnamefont{L.}~\bibnamefont{Campanelli}},\ }%
  \bibfield{journal}{%
  \Doi{10.1103/PhysRevD.70.083009}{\bibinfo {journal} {Phys.Rev.}}\ }%
  \textbf{\bibinfo {volume} {D70}},\ \bibinfo {pages} {083009} (\bibinfo {year}
  {2004}),\
  \Eprint{http://arxiv.org/abs/astro-ph/0407056}{arXiv:astro-ph/0407056
  [astro-ph]}%
  \bibAnnoteFile{NoStop}{Campanelli:2004wm}%
%%CITATION = ASTRO-PH/0407056;%%
\bibitem{Jedamzik:1994dd}%
  \BibitemOpen
  \bibfield{author}{%
  \bibinfo {author} {\bibfnamefont{K.}~\bibnamefont{Jedamzik}}\ and\ \bibinfo
  {author} {\bibfnamefont{G.~M.}\ \bibnamefont{Fuller}},\ }%
  \bibfield{journal}{%
  \Doi{10.1086/173788}{\bibinfo {journal} {Astrophys.J.}}\ }%
  \textbf{\bibinfo {volume} {423}},\ \bibinfo {pages} {33} (\bibinfo {year}
  {1994}),\
  \Eprint{http://arxiv.org/abs/astro-ph/9312063}{arXiv:astro-ph/9312063
  [astro-ph]}%
  \bibAnnoteFile{NoStop}{Jedamzik:1994dd}%
%%CITATION = ASTRO-PH/9312063;%%
\bibitem{Planck_param}%
  \BibitemOpen
  \bibfield{author}{%
  \bibinfo {author} {\bibfnamefont{P.}~\bibnamefont{Ade}} \emph{et~al.}
  (\bibinfo {collaboration} {Planck Collaboration})}%
   (\bibinfo {year} {2013}),\
  \Eprint{http://arxiv.org/abs/1303.5076}{arXiv:1303.5076 [astro-ph.CO]}%
  \bibAnnoteFile{NoStop}{Planck_param}%
%%CITATION = ARXIV:1303.5076;%%
\bibitem{Jedamzik:1996wp}%
  \BibitemOpen
  \bibfield{author}{%
  \bibinfo {author} {\bibfnamefont{K.}~\bibnamefont{Jedamzik}}, \bibinfo
  {author} {\bibfnamefont{V.}~\bibnamefont{Katalinic}},\ and\ \bibinfo {author}
  {\bibfnamefont{A.~V.}\ \bibnamefont{Olinto}},\ }%
  \bibfield{journal}{%
  \Doi{10.1103/PhysRevD.57.3264}{\bibinfo {journal} {Phys.Rev.}}\ }%
  \textbf{\bibinfo {volume} {D57}},\ \bibinfo {pages} {3264} (\bibinfo {year}
  {1998}),\
  \Eprint{http://arxiv.org/abs/astro-ph/9606080}{arXiv:astro-ph/9606080
  [astro-ph]}%
  \bibAnnoteFile{NoStop}{Jedamzik:1996wp}%
%%CITATION = ASTRO-PH/9606080;%%
\bibitem{Planck2015}%
  \BibitemOpen
  \bibfield{author}{%
  \bibinfo {author} {\bibfnamefont{P.~A.~R.}\ \bibnamefont{Ade}} \emph{et~al.}
  (\bibinfo {collaboration} {Planck})}%
   (\bibinfo {year} {2015}),\
  \Eprint{http://arxiv.org/abs/1502.01589}{arXiv:1502.01589 [astro-ph.CO]}%
  \bibAnnoteFile{NoStop}{Planck2015}%
%%CITATION = ARXIV:1502.01589;%%
\bibitem{PIXIE}%
  \BibitemOpen
  \bibfield{author}{%
  \bibinfo {author} {\bibfnamefont{A.}~\bibnamefont{Kogut}}, \bibinfo {author}
  {\bibfnamefont{D.}~\bibnamefont{Fixsen}}, \bibinfo {author}
  {\bibfnamefont{D.}~\bibnamefont{Chuss}}, \bibinfo {author}
  {\bibfnamefont{J.}~\bibnamefont{Dotson}}, \bibinfo {author}
  {\bibfnamefont{E.}~\bibnamefont{Dwek}}, \emph{et~al.},\ }%
  \bibfield{journal}{%
  \Doi{10.1088/1475-7516/2011/07/025}{\bibinfo {journal} {JCAP}}\ }%
  \textbf{\bibinfo {volume} {1107}},\ \bibinfo {pages} {025} (\bibinfo {year}
  {2011}),\ \Eprint{http://arxiv.org/abs/1105.2044}{arXiv:1105.2044
  [astro-ph.CO]}%
  \bibAnnoteFile{NoStop}{PIXIE}%
%%CITATION = ARXIV:1105.2044;%%
\bibitem{Planck2015Mag}%
  \BibitemOpen
  \bibfield{author}{%
  \bibinfo {author} {\bibfnamefont{P.~A.~R.}\ \bibnamefont{Ade}} \emph{et~al.}
  (\bibinfo {collaboration} {Planck})}%
   (\bibinfo {year} {2015}),\
  \Eprint{http://arxiv.org/abs/1502.01594}{arXiv:1502.01594 [astro-ph.CO]}%
  \bibAnnoteFile{NoStop}{Planck2015Mag}%
%%CITATION = ARXIV:1502.01594;%%
\bibitem{Aoki:2006we}%
  \BibitemOpen
  \bibfield{author}{%
  \bibinfo {author} {\bibfnamefont{Y.}~\bibnamefont{Aoki}}, \bibinfo {author}
  {\bibfnamefont{G.}~\bibnamefont{Endrodi}}, \bibinfo {author}
  {\bibfnamefont{Z.}~\bibnamefont{Fodor}}, \bibinfo {author}
  {\bibfnamefont{S.}~\bibnamefont{Katz}},\ and\ \bibinfo {author}
  {\bibfnamefont{K.}~\bibnamefont{Szabo}},\ }%
  \bibfield{journal}{%
  \Doi{10.1038/nature05120}{\bibinfo {journal} {Nature}}\ }%
  \textbf{\bibinfo {volume} {443}},\ \bibinfo {pages} {675} (\bibinfo {year}
  {2006}),\ \Eprint{http://arxiv.org/abs/hep-lat/0611014}{arXiv:hep-lat/0611014
  [hep-lat]}%
  \bibAnnoteFile{NoStop}{Aoki:2006we}%
%%CITATION = HEP-LAT/0611014;%%
\bibitem{Laine:1998qk}%
  \BibitemOpen
  \bibfield{author}{%
  \bibinfo {author} {\bibfnamefont{M.}~\bibnamefont{Laine}}\ and\ \bibinfo
  {author} {\bibfnamefont{K.}~\bibnamefont{Rummukainen}},\ }%
  \bibfield{journal}{%
  \Doi{10.1016/S0550-3213(98)00530-6}{\bibinfo {journal} {Nucl.Phys.}}\ }%
  \textbf{\bibinfo {volume} {B535}},\ \bibinfo {pages} {423} (\bibinfo {year}
  {1998}),\ \Eprint{http://arxiv.org/abs/hep-lat/9804019}{arXiv:hep-lat/9804019
  [hep-lat]}%
  \bibAnnoteFile{NoStop}{Laine:1998qk}%
%%CITATION = HEP-LAT/9804019;%%
\bibitem{Vachaspati:2001nb}%
  \BibitemOpen
  \bibfield{author}{%
  \bibinfo {author} {\bibfnamefont{T.}~\bibnamefont{Vachaspati}},\ }%
  \bibfield{journal}{%
  \Doi{10.1103/PhysRevLett.87.251302}{\bibinfo {journal} {Phys.Rev.Lett.}}\ }%
  \textbf{\bibinfo {volume} {87}},\ \bibinfo {pages} {251302} (\bibinfo {year}
  {2001}),\
  \Eprint{http://arxiv.org/abs/astro-ph/0101261}{arXiv:astro-ph/0101261
  [astro-ph]}%
  \bibAnnoteFile{NoStop}{Vachaspati:2001nb}%
%%CITATION = ASTRO-PH/0101261;%%
\bibitem{Wagstaff:2014fla}%
  \BibitemOpen
  \bibfield{author}{%
  \bibinfo {author} {\bibfnamefont{J.~M.}\ \bibnamefont{Wagstaff}}\ and\
  \bibinfo {author} {\bibfnamefont{R.}~\bibnamefont{Banerjee}}}%
   (\bibinfo {year} {2014}),\
  \Eprint{http://arxiv.org/abs/1409.4223}{arXiv:1409.4223 [astro-ph.CO]}%
  \bibAnnoteFile{NoStop}{Wagstaff:2014fla}%
%%CITATION = ARXIV:1409.4223;%%
\bibitem{Boyarsky:2011uy}%
  \BibitemOpen
  \bibfield{author}{%
  \bibinfo {author} {\bibfnamefont{A.}~\bibnamefont{Boyarsky}}, \bibinfo
  {author} {\bibfnamefont{J.}~\bibnamefont{Frohlich}},\ and\ \bibinfo {author}
  {\bibfnamefont{O.}~\bibnamefont{Ruchayskiy}},\ }%
  \bibfield{journal}{%
  \Doi{10.1103/PhysRevLett.108.031301}{\bibinfo {journal} {Phys. Rev. Lett.}}\
  }%
  \textbf{\bibinfo {volume} {108}},\ \bibinfo {pages} {031301} (\bibinfo {year}
  {2012}),\ \Eprint{http://arxiv.org/abs/1109.3350}{arXiv:1109.3350
  [astro-ph.CO]}%
  \bibAnnoteFile{NoStop}{Boyarsky:2011uy}%
%%CITATION = ARXIV:1109.3350;%%
\bibitem{Boyarsky:2015faa}%
  \BibitemOpen
  \bibfield{author}{%
  \bibinfo {author} {\bibfnamefont{A.}~\bibnamefont{Boyarsky}}, \bibinfo
  {author} {\bibfnamefont{J.}~\bibnamefont{Frohlich}},\ and\ \bibinfo {author}
  {\bibfnamefont{O.}~\bibnamefont{Ruchayskiy}}}%
   (\bibinfo {year} {2015}),\
  \Eprint{http://arxiv.org/abs/1504.04854}{arXiv:1504.04854 [hep-ph]}%
  \bibAnnoteFile{NoStop}{Boyarsky:2015faa}%
%%CITATION = ARXIV:1504.04854;%%
\bibitem{Tashiro:2013ita}%
  \BibitemOpen
  \bibfield{author}{%
  \bibinfo {author} {\bibfnamefont{H.}~\bibnamefont{Tashiro}}, \bibinfo
  {author} {\bibfnamefont{W.}~\bibnamefont{Chen}}, \bibinfo {author}
  {\bibfnamefont{F.}~\bibnamefont{Ferrer}},\ and\ \bibinfo {author}
  {\bibfnamefont{T.}~\bibnamefont{Vachaspati}}}%
   (\bibinfo {year} {2013}),\ \doi{\bibinfo {doi} {10.1093/mnrasl/slu134}},\
  \Eprint{http://arxiv.org/abs/1310.4826}{arXiv:1310.4826 [astro-ph.CO]}%
  \bibAnnoteFile{NoStop}{Tashiro:2013ita}%
%%CITATION = ARXIV:1310.4826;%%
\bibitem{Ganc:2014wia}%
  \BibitemOpen
  \bibfield{author}{%
  \bibinfo {author} {\bibfnamefont{J.}~\bibnamefont{Ganc}}\ and\ \bibinfo
  {author} {\bibfnamefont{M.~S.}\ \bibnamefont{Sloth}},\ }%
  \bibfield{journal}{%
  \Doi{10.1088/1475-7516/2014/08/018}{\bibinfo {journal} {JCAP}}\ }%
  \textbf{\bibinfo {volume} {1408}},\ \bibinfo {pages} {018} (\bibinfo {year}
  {2014}),\ \Eprint{http://arxiv.org/abs/1404.5957}{arXiv:1404.5957
  [astro-ph.CO]}%
  \bibAnnoteFile{NoStop}{Ganc:2014wia}%
%%CITATION = ARXIV:1404.5957;%%
\end{thebibliography}%

\end{document}